\documentclass[11pt,fleqn]{article}

\usepackage{amsmath,amssymb,amsthm,enumerate,cite} 

\newcommand{\p}{\partial}
\newcommand{\const}{\mathop{\rm const}\nolimits}
\newcommand{\Equiv}{\mathop{\sim}}
\newcommand{\CV}{\mathop{\rm CV}\nolimits}
\newcommand{\CL}{\mathop{\rm CL}\nolimits}
\newcommand{\Ch}{\mathop{\rm Ch}\nolimits}

\newcommand{\Eop}{\mathop{\sf E}}
\newcommand{\Fder}{\mathop{\sf D}}

\newtheorem{theorem}{Theorem}

\newtheorem{corollary}{Corollary}
\newtheorem{lemma}{Lemma}
{\theoremstyle{definition}
\newtheorem{definition}{Definition}
\newtheorem{note}{Note}
\newtheorem*{note*}{Note}
\newtheorem{proposition}{Proposition}
}

\begin{document}
\begin{center}
{\LARGE\bf Hierarchy of Conservation Laws \\ of
Diffusion--Convection Equations\\[1.3ex]}
{\large\bf Roman~O.~Popovych~$^\dag$ and Nataliya~M.~Ivanova~$^\ddag$
\\[1.7ex]}
\footnotesize
Institute of Mathematics of National Academy of Sciences of Ukraine, \\
3 Tereshchenkivska Str., Kyiv-4, 01601, Ukraine\\
\vspace{1em}
E-mail: $^\dag$rop@imath.kiev.ua, $^\ddag$ivanova@imath.kiev.ua
\end{center}

\begin{abstract}
We introduce notions of equivalence of conservation laws with respect to Lie symmetry groups
for fixed systems of differential equations and with respect to equivalence groups
or sets of admissible transformations for classes of such systems.
We also revise the notion of linear dependence of conservation laws and
define the notion of local dependence of potentials.
To construct conservation laws, we develop and apply the most direct method which is effective to use
in the case of two independent variables.
Admitting possibility of dependence of conserved vectors on a number of potentials, we generalize
the iteration procedure proposed by Bluman and \mbox{Doran-Wu} for finding nonlocal (potential) conservation laws.
As~an~example, we completely classify potential conservation laws (including arbitrary order local ones)
of diffusion--convection equations with respect to the equivalence group
and construct an exhaustive list of locally inequivalent potential systems corresponding to these equations.

\vspace{2ex}

\end{abstract}

\section{Introduction}

After the Emmy Noether's remarkable paper~\cite{Noether1918} had become well-known,
a number of authors (see e.g.~\cite{Olver1986,Ibragimov&Kara&Mahomed1998,Steinberg&Wolf1981,Strampp1982a})
searched for conservation laws using the symmetry approach based on the Noether's results.
In view of the generalized Noether's theorem~\cite{Olver1986}, there exists one-to-one correspondence between
the non-trivial generalized variational symmetries of some functional and the non-trivial conservation laws
of the associated Euler--Lagrange equations,
and any such symmetry is a generalized symmetry of the Euler--Lagrange equations.

The Noether's approach has a number of advantages.
It reduces construction of conservation laws to finding symmetries
for which there exist a number of well-developed methods, and complete description
of necessary symmetry properties is known for a lot of systems of differential equations.
However, this approach can be applied only to Euler--Lagrange equations that form normal systems and admit
symmetry groups satisfying an additional ``variational'' property of leaving the variational integral invariant in
some sense~\cite{Olver1986}. The latter requirements lead to restriction of class of systems that could be
investigated in such way.

At~the same time, the definition of conservation laws itself gives rise to a method of finding conservation laws.
Technique of calculations used in the framework of this method is similar to
the classical Lie method yielding symmetries of differential equations~\cite[Chapter~6]{Ibragimov1994V1}.
As~mentioned in the above reference, such algorithmic possibility was first employed by
\mbox{P.-S.}~Laplace~\cite{Laplace1798}
for derivation of the well-known Laplace vector of the two-body Kepler problem.
Following tradition from group analysis of differential equations,
we may call this method {\em direct} and distinguish four its versions,
depending on the way of taking into account systems under investigation.
(See e.g.~\cite{Bluman&Doran-Wu1995,Anco&Bluman2002a,Anco&Bluman2002b,Kara&Mahomed2000}
and Section~\ref{SectionDirectIterationMethod} of this paper for more details
as well as~\cite{Wolf2002} for comparison of the versions and their realizations in computer algebra programs.)
In the present paper we use the most direct version based on immediate solving of determining equations for
conserved vectors of conservations laws on the solution manifolds of investigated systems.

Let us note that there exist other approaches for construction of conservation laws
which differ from the Noether's or above direct ones,
are based on exploitation of symmetry properties of differential equations
and can be applied  to non-Lagrangian systems.
Thus, W.I.~Fushchych and A.G.~Nikitin~\cite{Fushchych&Nikitin1994} proposed to calculate directly bilinear combinations
of solutions of motion equations, which are conserved in time by virtue of symmetries of these equations.
It is possible in such way to find conservation laws corresponding to non-geometric symmetries.
In the recent work~\cite{Anco2003} a purely algebraic formula has been derived for generating conservation laws
of systems of differential equations that possess a scaling-invariance.

To classify conservation laws, instead of the usual equivalence relation on their set
(more exactly, on the set of conserved vectors)
we propose to use the natural and more general notions of equivalences of conservation laws with respect to
Lie symmetry groups for fixed systems of differential equations
and with respect to equivalence groups or sets of admissible point (or contact) transformations
for classes of such systems.
Results of classification up to these equivalences
are more comprehensible, especially, if a whole class of systems is studied and
blend with the framework of group analysis.

G.~Bluman and P.R.~Doran-Wu~\cite{Bluman&Doran-Wu1995} proposed
an ingenious procedure of branching iterations for finding nonlocal (potential) conservation laws
of diffusion equations.
Namely, on each iteration they use a conservation law from the previous iteration
(one conservation law for one iteration) to introduce a potential and to construct the
extended potential system. Then they study local conservation laws of the potential system,
which are, generally speaking, nonlocal (potential) conservation laws for the initial equation.
To the best of our knowledge, it was the first paper where the idea of hierarchy of potential systems and
associated conservation laws is presented in an explicit form.

We generalize the iteration procedure by G.~Bluman and P.~Doran-Wu,
admitting dependence of conserved vectors on different number (from one to the maximum possible that)
of new potentials on each iteration.
The idea of a similar approach was adduced in~\cite{Wahlquist&Estabrook1975} and
was formalized in the form of notion of {\em universal Abelian covering}
of differential equations~\cite{Vinogradov1984,Bocharov&Co1997,Sergyeyev2000}.
Such approach naturally results in the questions on some independence of employed potentials.
That is why, in this paper we consider definition of linear dependence of conservation laws in detail
and define the notion of local dependence of potentials.

\looseness=-1
Therefore, in the first part of the paper we propose some technique and
discuss the classification problem for conservation laws in general.
The ultimate goal of the second part is to present an exhaustive classification of potential conservation laws
in a quite difficult and interesting case.
As an illustration of the proposed technique, we choose the class of diffusion--convection equations
\begin{equation} \label{eqf1}
u_t=(A(u)u_x)_x+B(u)u_x,
\end{equation}
where $A=A(u)$ and $B=B(u)$ are arbitrary smooth functions of $u$, $A(u)\!\neq\! 0$.
Symmetry properties of~\eqref{eqf1} were considered in~\cite{Edwards,Oron&Rosenau,Yung&Verburg&Baveye1994}, however
the complete and strong group classification of~\eqref{eqf1} was first presented
in~\cite{Popovych&Ivanova2004NVCDCEs} (see also references therein for more
details about symmetry analysis of classes intersecting class~\eqref{eqf1}).

Studying conservation laws of equations~\eqref{eqf1} was started from linear equations~\cite{Steinberg&Wolf1981}.
V.A.~Dorodnitsyn and S.R.~Svirshchevskii~\cite{Dorodnitsyn&Svirshchevskii1983}
(see also~\cite[Chapter~10]{Ibragimov1994V1})
completely investigated the local conservation laws for reaction--diffusion equations $u_t=(A(u)u_x)_x+C(u)$.
The~first-order local conservation laws of equations~\eqref{eqf1} were constructed by
A.H.~Kara and F.M.~Mahomed~\cite{Kara&Mahomed2002}.
Developing results of~\cite{Bluman&Doran-Wu1995} obtained for the case~$B=0$,
N.M.~Ivanova~\cite{Ivanova2004} classified
the first-order local conservation laws for equations~\eqref{eqf1} with respect
to the equivalence group and constructed potential conservation laws of the first level.
Namely, she made two steps of the iteration procedure, looking, in the second step,
for the first-order local conservation laws of the potential systems obtained after the first step.

In the present work we exhaustively classify,
with respect to the corresponding equivalence group, the local conservation laws of an arbitrary order,
find the simplest and general potential conservation laws and construct locally inequivalent potential systems
of equations~\eqref{eqf1}.
All~possible steps of the branching iteration procedure are done,
and admission of dependence of conserved vectors on a number of potentials is of fundamental importance
for completing iterations.
We obtain eight inequivalent cases of equations~\eqref{eqf1} having different sets of potential conservation laws:
\begin{itemize}
\item
the general case (the parameter-functions~$A$ and~$B$ are arbitrary);
\item
three cases with an arbitrary value of~$A$ and a special value of $B$ ($B=0$; $B=A$; $B=\int\! A\,du+uA$);
\item
three corresponding linearizable equations ($A=u^{-2}$, $B=0$; $A=B=u^{-2}$; $A=1$, $B=2u$) and
\item
the linear heat equation ($A=1$, $B=0$).
\end{itemize}

The latter case takes on special significance in our consideration
since investigation of the linearizable equations having infinite series
of potential conservation laws is reduced to this case
and the non-linearizable equations from class~\eqref{eqf1} have at the most two independent conservation~laws.

Our paper is organized as follows.
First of all (Section~\ref{SectionBasicDef}) we give a basic theoretical background,
following the spirit of~\cite{Olver1986}.
We recall the notions of equivalence of conserved vectors and characteristics with respect to the
triviality relation and discuss properties of the space of conserved vectors
and the space of characteristics.
This naturally results in the notions of linear dependence of conservation laws and local dependence of potentials
(Sections~\ref{SectionDirectIterationMethod} and~\ref{Section2DimCase}).

In~Section~\ref{SectionEquivOfConsLaws}
we introduce the notions of equivalence of conservation laws with respect
to Lie symmetry groups for fixed systems of differential equations and equivalence groups
or sets of admissible point (or contact) transformations for classes of such systems.
We emphasize possibility of solving classification problems for conservation laws with respect to the above
equivalences similarly to usual group classification problems for differential equations.

In~Section~\ref{SectionDirectIterationMethod} we adduce different versions of the direct method
of construction of conservation laws, emphasizing the most direct one and combining it with classification
up to symmetry or equivalence groups.
We also generalize the iteration method of finding potential conservation laws and associated potential systems.

Since the two-dimensional case is special for construction of potential systems
we describe it in more detail in Section~\ref{Section2DimCase}.

Then we apply the theoretical background given in the previous sections to investigation of
diffusion--convection equations from class~\eqref{eqf1}.
The local conservation laws of~\eqref{eqf1} are classified
with respect to the corresponding equivalence group in Section~\ref{SectionLocalConsLwsOfDCEs}.
In Section~\ref{SectionSimplestPotentialConservationLaws} we construct simplest potential conservation laws
and analyze connections between them using potential equivalence transformations.
The subject of Section~\ref{SectionPotConsLawLHE} is the description of general potential conservation laws
of the linear heat equation.
In Section~\ref{SectionGenPotConsLaws} we complete studying potential conservation laws of
diffusion--convection equations
and adduce the hierarchy of conservation laws obtained in our framework.
In the same section we give an exhaustive list of locally inequivalent potential systems of equations~\eqref{eqf1}.
The obtained results can be interpreted as construction of universal Abelian covering for the whole class of
diffusion--convection equations.

\pagebreak

\section{Basic definitions and statements}\label{SectionBasicDef}

Let~$\mathcal{L}$ be a system~$L(x,u_{(\rho)})=0$ of $l$ differential equations $L^1=0$, \ldots, $L^l=0$
for $m$ unknown functions $u=(u^1,\ldots,u^m)$
of $n$ independent variables $x=(x_1,\ldots,x_n).$
Here $u_{(\rho)}$ denotes the set of all the derivatives of the functions $u$ with respect to $x$
of order no greater than~$\rho$, including $u$ as the derivatives of the zero order.
Let $\mathcal{L}_{(k)}$ denote the set of all algebraically independent differential consequences
that have, as differential equations, orders no greater than $k$. We identify~$\mathcal{L}_{(k)}$ with
the manifold determined by~$\mathcal{L}_{(k)}$ in the jet space~$J^{(k)}$.

\begin{definition}\label{def.conservation.law}
A {\em conservation law} of the system~$\mathcal{L}$ is a divergence expression
${\rm Div}\,F:=D_iF^i$ which vanishes for all solutions of~$\mathcal{L}$: ${\rm Div}F\bigl|_\mathcal{L}=0.$
The $n$-tuple $F=(F^1(x,u_{(r)}),\ldots,F^n(x,u_{(r)}))$ is called a {\em conserved vector}
of this (local) conservation law.
\end{definition}

In Definition~\ref{def.conservation.law} and below
$D_i=D_{x_i}$ denotes the operator of total differentiation with respect to the variable~$x_i$, i.e.
$D_i=\p_{x_i}+u^a_{\alpha,i}\p_{u^a_\alpha}$, where
$u^a_\alpha$ and $u^a_{\alpha,i}$ stand for the variables in jet spaces,
which correspond to derivatives
$\p^{|\alpha|}u^a/\p x_1^{\alpha_1}\ldots\p x_n^{\alpha_n}$ and $\p u^a_\alpha/\p x_i$,
$\alpha=(\alpha_1,\ldots,\alpha_n)$,
$\alpha_i\in\mathbb{N}\cup\{0\}$, $|\alpha|{:}=\alpha_1+\cdots+\alpha_n$.
We use the summation convention for repeated indices and assume any function as its zero-order derivative.
The notation~$V\bigl|_\mathcal{L}$ means that values of $V$ are considered
only on solutions of the system~$\mathcal{L}$.

\begin{definition}
A conserved vector $F$ is called {\em trivial} if
$
F^i=\hat F^i+\check F^i, \quad i=\overline{1,n},
$
where $\hat F^i$ and $\check F^i$ are, likewise $F^i$, functions of $x$ and derivatives of $u$
(i.e. differential functions),
$\hat F^i$ vanish on the solutions of~$\mathcal L$ and the $n$-tuple $\check F=(\check F^1,\ldots,\check F^n)$
is a null divergence (i.e. its divergence vanishes identically).
\end{definition}

The triviality concerning the vanishing conserved vectors on solutions of the system
can be easily
eliminated by confining on the manifold of the system, taking into account all its necessary differential consequences.

A characterization of all null divergences is given by the following lemma (see e.g.~\cite{Olver1986}).

\begin{lemma}\label{lemma.null.divergence}
The $n$-tuple $F=(F^1,\ldots,F^n)$, $n\ge2$, is a null divergence ($\mathop{\rm Div}\nolimits F\equiv0$)
iff there exist smooth functions $v^{ij}$ ($i,j=\overline{1,n}$) of $x$ and derivatives of $u$,
such that $v^{ij}=-v^{ji}$ and $F^i=D_jv^{ij}$.
\end{lemma}

The functions $v^{ij}$ are called {\em potentials} corresponding to the null divergence~$F$.
If $n=1$ any null divergence is constant.

\begin{definition}\label{def.conservation.law.equivalence}
Two conserved vectors $F$ and $F'$ are called {\em equivalent} if
the vector-function $F'-F$ is a trivial conserved vector.
\end{definition}

\begin{note}\label{NoteFushchychNikitin1992}
Sometimes other definitions of equivalence of conservation laws are used~\cite{Fushchych&Nikitin1992}.
\end{note}

By the latter definition any trivial conserved vector is equivalent to the vanishing one.
For any system~$\mathcal{L}$ of differential equations the set~$\CV(\mathcal{L})$ of conserved vectors of
its conservation laws is a linear space,
and the subset~$\CV_0(\mathcal{L})$ of trivial conserved vectors is a linear subspace in~$\CV(\mathcal{L})$.
The factor space~$\CL(\mathcal{L})=\CV(\mathcal{L})/\CV_0(\mathcal{L})$
coincides with the set of equivalence classes of~$\CV(\mathcal{L})$ with respect to the equivalence relation adduced in
Definition~\ref{def.conservation.law.equivalence}.
We can identify elements of~$\CL(\mathcal{L})$ with conservation laws
and call~$\CL(\mathcal{L})$ also as {\em the space of conservation laws} of~$\mathcal{L}$
(see e.g.~\cite{Zharinov1986}).
That is why we assume description of the set of conservation laws
as finding~$\CL(\mathcal{L})$ that is equivalent to construction of either a basis if
$\dim \CL(\mathcal{L})<\infty$ or a system of generatrices in the infinite dimensional case,
and we will additionally identify elements from~$\CL(\mathcal{L})$ with their representatives
in~$\CV(\mathcal{L})$.
In contrast to the order $r_F$ of a conserved vector~$F$ as the maximal order of derivatives explicitly appearing in~$F$,
the {\em order of a conservation law} as an element~$\cal F$ from~$\CL(\mathcal{L})$
is called $\min\{r_F\,|\,F\in\CV(\mathcal{L})\ \mbox{corresponds to}\ {\cal F}\}$.
Under linear dependence of conservation laws we understand linear dependence of corresponding elements
in~$\CL(\mathcal{L})$.

\begin{definition}\label{def.conservation.law.dependence}
Conservation laws of a system~$\mathcal{L}$ are called {\em linearly dependent} if
there exists their linear combination having a trivial conserved vector.
\end{definition}


Let the system~$\cal L$ be totally nondegenerate~\cite{Olver1986}.
Then application of the Hadamard lemma to the definition of conservation law and integrating by parts imply that
the left side of any conservation law of~$\mathcal L$ can be always presented up to the equivalence relation
as a linear combination of left side of independent equations from $\mathcal L$
with coefficients~$\lambda^\mu$ being functions on a suitable jet space~$J^{(k)}$:
\begin{equation}\label{CharFormOfConsLaw}
\mathop{\rm Div}\nolimits F=\lambda^\mu L^\mu.
\end{equation}
Here the order~$k$ is determined by~$\mathcal L$ and the allowable order of conservation laws,
$\mu=\overline{1,l}$.

\begin{definition}\label{DefCharForm}
Formula~\eqref{CharFormOfConsLaw} and the $l$-tuple $\lambda=(\lambda^1,\ldots,\lambda^l)$
are called the {\it characteristic form} and the {\it characteristic}
of the conservation law~$\mathop{\rm Div}\nolimits F=0$ correspondingly.
\end{definition}

The characteristic~$\lambda$ is {\em trivial} if it vanishes for all solutions of $\cal L$.
Since $\cal L$ is nondegenerate, the characteristics~$\lambda$ and~$\tilde\lambda$ satisfy~\eqref{CharFormOfConsLaw}
for the same~$F$ and, therefore, are called {\em equivalent}
iff $\lambda-\tilde\lambda$ is a trivial characteristic.
Similarly to conserved vectors, the set~$\Ch(\mathcal{L})$ of characteristics
corresponding to conservation laws of the system~$\cal L$ is a linear space,
and the subset~$\Ch_0(\mathcal{L})$ of trivial characteristics is a linear subspace in~$\Ch(\mathcal{L})$.
The factor space~$\Ch_{\rm f}(\mathcal{L})=\Ch(\mathcal{L})/\Ch_0(\mathcal{L})$
coincides with the set of equivalence classes of~$\Ch(\mathcal{L})$
with respect to the above characteristic equivalence relation.

The following result~\cite{Olver1986} forms the cornerstone for the methods of studying conservation laws,
which are based on formula~\eqref{CharFormOfConsLaw}, including the Noether theorem and the direct method by
Anco and Bluman~\cite{Anco&Bluman2002a,Anco&Bluman2002b}.

\begin{theorem}[\cite{Olver1986}]\label{TheoremIsomorphismChCV}
Let~$\mathcal{L}$ be a normal, totally nondegenerate system of differential equations.
Then representation of conservation laws of~$\mathcal{L}$ in the characteristic form~\eqref{CharFormOfConsLaw}
generates a one-to-one linear mapping between~$\CL(\mathcal{L})$ and~$\Ch_{\rm f}(\mathcal{L})$.
\end{theorem}

Using properties of total divergences, we can exclude the conserved vector~$F$ from~\eqref{CharFormOfConsLaw}
and obtain a condition for the characteristic~$\lambda$ only.
Namely, a differential function~$f$ is a total divergence, i.e. $f=\mathop{\rm Div} F$
for some $n$-tuple~$F$ of differential functions iff $\Eop(f)=0$.
Hereafter the Euler operator~$\Eop=(\Eop^1,\ldots, \Eop^m)$ is the $m$-tuple of differential operators
\[
{\Eop}^a=(-D)^\alpha\p_{u^a_\alpha}, \quad a=\overline{1,m},
\]
where
$\alpha=(\alpha_1,\ldots,\alpha_n)$ runs the multi-indices set ($\alpha_i\!\in\!\mathbb{N}\cup\{0\}$),
$(-D)^\alpha=(-D_1)^{\alpha_1}\ldots(-D_m)^{\alpha_m}$.
Therefore, action of the Euler operator on~\eqref{CharFormOfConsLaw}
results to the equation
\begin{equation}\label{NSCondOnChar}
\Eop(\lambda^\mu L^\mu)={\Fder}_\lambda^*(L)+{\Fder}_L^*(\lambda)=0,
\end{equation}
which is a necessary and sufficient condition on characteristics of conservation laws for the system~$\mathcal{L}$.
The matrix differential operators~${\Fder}_\lambda^*$ and~${\Fder}_L^*$ are the adjoints of
the Fr\'echet derivatives~${\Fder}_\lambda^{\phantom{*}}$ and~${\Fder}_L^{\phantom{*}}$, i.e.
\[
{\Fder}_\lambda^*(L)=\left((-D)^\alpha\left( \dfrac{\p\lambda^\mu}{\p u^a_\alpha}L^\mu\right)\right), \qquad
{\Fder}_L^*(\lambda)=\left((-D)^\alpha\left( \dfrac{\p L^\mu}{\p u^a_\alpha}\lambda^\mu\right)\right).
\]
Since ${\Fder}_\lambda^*(L)=0$ automatically on solutions of~$\mathcal{L}$ then
equation~\eqref{NSCondOnChar} implies a necessary condition for $\lambda$ to belong to~$\Ch(\mathcal{L})$:
\begin{equation}\label{NCondOnChar}
{\Fder}_L^*(\lambda)\bigl|_{\mathcal{L}}=0.
\end{equation}
Condition~\eqref{NCondOnChar} can be considered as adjoint to the criteria
${\Fder}_L^{\phantom{*}}(\eta)\bigl|_{\mathcal{L}}=0$ for infinitesimal invariance of $\mathcal{L}$
with respect to evolutionary vector field having the characteristic~$\eta=(\eta^1,\ldots,\eta^m)$.
That is why solutions of~\eqref{NCondOnChar} are called sometimes as
{\em cosymmetries}~\cite{Sergyeyev2000,Blaszak1998} or
{\em adjoint symmetries}~\cite{Anco&Bluman2002b}.

\section{Equivalence of conservation laws}\label{SectionEquivOfConsLaws}


We can essentially simplify and order classification of conservation laws, taking into account additionally
symmetry transformations of a system or equivalence transformations of a whole class of systems.
Such problem is similar to one of group classification of differential equations.

\begin{proposition}
Any point transformation~$g$ maps a class of equations in the conserved form into itself.
More exactly, the transformation~$g$: $\tilde x=x_g(x,u)$, $\tilde u=u_g(x,u)$ prolonged to the jet space~$J^{(r)}$
transforms the equation $D_iF^i=0$ to the equation $D_iF^i_g=0$. The transformed conserved vector~$F_g$ is determined
by the formula
\begin{equation}\label{eq.tr.var.cons.law}
F_g^i(\tilde x,\tilde u_{(r)})=\frac{D_{x_j}\tilde x_i}{|D_x\tilde x|}\,F^j(x,u_{(r)}),
\quad\mbox{i.e.}\quad
F_g(\tilde x,\tilde u_{(r)})=\frac{1}{|D_x\tilde x|}(D_x\tilde x)F(x,u_{(r)})
\end{equation}
in the matrix notions. Here $|D_x\tilde x|$ is the determinant of the matrix $D_x\tilde x=(D_{x_j}\tilde x_i)$.
\end{proposition}

\begin{note}
In the case of one dependent variable ($m=1$) $g$ can be a contact transformation:
$\tilde x=x_g(x,u_{(1)})$, $\tilde u_{(1)}=u_{g(1)}(x,u_{(1)})$.
Similar notes are also true for the statements below.
\end{note}

\begin{definition}
Let $G$ be a symmetry group of the system~$\mathcal{L}$.
Two conservation laws with the conserved vectors $F$ and $F'$ are called {\em $G$-equivalent} if
there exists a transformation $g\in G$ such that the conserved vectors $F_g$ and $F'$
are equivalent in the sense of Definition~\ref{def.conservation.law.equivalence}.
\end{definition}

Any transformation $g\in G$ induces a linear one-to-one mapping $g_*$ in~$\CV(\mathcal{L})$,
transforms trivial conserved vectors only to trivial ones
(i.e. $\CV_0(\mathcal{L})$ is invariant with respect to~$g_*$)
and therefore induces a linear one-to-one mapping $g_{\rm f}$ in~$\CL(\mathcal{L})$.
It is obvious that $g_{\rm f}$ preserves linear (in)dependence of elements
in~$\CL(\mathcal{L})$ and maps a basis (a set of generatrices) of~$\CL(\mathcal{L})$
in a basis (a set of generatrices) of the same space.
In such way we can consider the $G$-equivalence relation of conservation laws
as well-determined on~$\CL(\mathcal{L})$ and use it to classify conservation laws.

\begin{proposition}
If the system~$\mathcal{L}$ admits a one-parameter group of transformations then the infinitesimal generator
$X=\xi^i\p_i+\eta^a\p_{u^a}$
of this group can be used for construction of new conservation laws from known ones.
Namely, differentiating equation~(\ref{eq.tr.var.cons.law})
with respect to the parameter $\varepsilon$ and taking the value $\varepsilon=0$,
we obtain the new conserved vector
\begin{equation}\label{eq.inf.tr.var.cons.law}
\widetilde F^i=-X_{(r)}F^i+(D_j\xi^i)F^j-(D_j\xi^j)F^i.
\end{equation}
Here $X_{(r)}$ denotes the $r$-th prolongation~\cite{Olver1986,Ovsiannikov1982} of the operator $X$.
\end{proposition}

\begin{note}Formula~\eqref{eq.inf.tr.var.cons.law} can be directly extended to generalized symmetry operators
(see, for example,~\cite{Anco&Bluman2002c,Kara&Mahomed2002}).
A similar statement for generalized symmetry operators in evolutionary form ($\xi^i=0$)
was known earlier~\cite{Ibragimov1985,Olver1986}.
It was used in~\cite{Khamitova1982} to introduce a notion of basis of conservation laws as a set
which generates a whole set of conservation laws with action of generalized symmetry operators and operation
of linear combination.
\end{note}

\begin{proposition}\label{PropositionOnInducedMapping}
Any point transformation $g$ between systems~$\mathcal{L}$ and~$\tilde{\mathcal{L}}$
induces a linear one-to-one mapping $g_*$ from~$\CV(\mathcal{L})$ into~$\CV(\tilde{\mathcal{L}})$,
which maps $\CV_0(\mathcal{L})$ into~$\CV_0(\tilde{\mathcal{L}})$
and generates a linear one-to-one mapping $g_{\rm f}$ from~$\CL(\mathcal{L})$ into~$\CL(\tilde{\mathcal{L}})$.
\end{proposition}

\begin{corollary}\label{CorollaryOnInducedMappingOfChar}
Any point transformation $g$ between systems~$\mathcal{L}$ and~$\tilde{\mathcal{L}}$
induces a linear one-to-one mapping $\hat g_{\rm f}$ from~$\Ch_{\rm f}(\mathcal{L})$
into~$\Ch_{\rm f}(\tilde{\mathcal{L}})$.
\end{corollary}
It is possible to obtain an explicit formula for correspondence between characteristics of~$\mathcal{L}$
and~$\tilde{\mathcal{L}}$.
Let $\tilde{\mathcal{L}}^\mu=\Lambda^{\mu\nu}\mathcal{L}^\nu$,
where $\Lambda^{\mu\nu}=\Lambda^{\mu\nu\alpha}D^\alpha$, $\Lambda^{\mu\nu\alpha}$ are differential functions,
$\alpha=(\alpha_1,\ldots,\alpha_n)$ runs the multi-indices set ($\alpha_i\!\in\!\mathbb{N}\cup\{0\}$),
$\mu,\nu=\overline{1,l}$.
Then
\[\lambda^\mu={\Lambda^{\nu\mu}}^*(|D_x\tilde x|\tilde\lambda^\nu).\]
Here ${\Lambda^{\nu\mu}}^*=(-D)^\alpha\cdot\Lambda^{\mu\nu\alpha}$ is the adjoint to the operator~$\Lambda^{\nu\mu}$.
For a number of cases, e.g. if~$\mathcal{L}$ and~$\tilde{\mathcal{L}}$ are single partial differential equations
($l=1$), the operators~$\Lambda^{\mu\nu}$ are simply differential functions
(i.e. $\Lambda^{\mu\nu\alpha}=0$ for $|\alpha|>0$) and, therefore, ${\Lambda^{\nu\mu}}^*=\Lambda^{\mu\nu}$.

Consider the class~$\mathcal{L}|_S$ of systems~$L(x,u_{(\rho)},\theta(x,u_{(\rho)}))=0$
parameterized with the parameter-functions~$\theta=\theta(x,u_{(\rho)}).$
Here $L$ is a tuple of fixed functions of $x,$ $u_{(\rho)}$ and $\theta.$
$\theta$ denotes the tuple of arbitrary (parametric) functions
$\theta(x,u_{(\rho)})=(\theta^1(x,u_{(\rho)}),\ldots,\theta^k(x,u_{(\rho)}))$
satisfying the condition~$S(x,u_{(\rho)},\theta_{(q)}(x,u_{(\rho)}))=0$.
This condition consists of differential equations on $\theta$,
where $x$ and $u_{(\rho)}$ play the role of independent variables
and $\theta_{(q)}$ stands for the set of all the partial derivatives of $\theta$ of order no greater than $q$.
In what follows we call the functions $\theta$ arbitrary elements.
Denote the local transformations group preserving the
form of systems from~$\mathcal{L}|_S$ as $G^{\Equiv}=G^{\Equiv}(L,S).$

Consider the set~$P=P(L,S)$ of all pairs each of which consists of
a system from~$\mathcal{L}|_S$ and a conservation law of this system.
In view of Proposition~\ref{PropositionOnInducedMapping},
action of transformations from~$G^{\Equiv}$ together with the pure equivalence relation of conserved vectors
naturally generates an equivalence relation on~$P$.
Classification of conservation laws with respect to~$G^{\Equiv}$ will be understood as
classification in~$P$ with respect to the above equivalence relation.
This problem can be investigated in the way that is similar to group classification in classes
of systems of differential equations. Namely, we construct firstly the conservation laws
that are defined for all values of the arbitrary elements.
(The corresponding conserved vectors may depend on the arbitrary elements.)
Then we classify, with respect to the equivalence group, arbitrary elements for each of that the system
admits additional conservation laws.

In an analogues way we also can introduce an equivalence relation on~$P$
generated by all admissible point or contact transformations
(called also form-preserving ones~\cite{Kingston&Sophocleous1998})
in pairs of equations from~$\mathcal{L}|_S$.

\begin{note}
It can be easy shown that all the above equivalences are indeed equivalence relations.
\end{note}

\section{Direct iteration method of finding conservation laws}\label{SectionDirectIterationMethod}

To construct conservation laws of a system~$\mathcal{L}$ of differential equations,
we iterate a modification of the most {\em direct method} based on Definition~\ref{def.conservation.law}.
More precisely, the algorithm is as follows.

\medskip

\noindent {\em Zeroth iteration.}
At first we construct local conservation laws of~$\mathcal{L}$.
We fix an (arbitrary) order~$r$ of conserved vectors under consideration.
Then we introduce local coordinates (``unconstrained variables'') on the manifold~${\mathcal L}_{(r+1)}$
determined by~the system~$\mathcal L$ and its differential consequences in~$J^{(r+1)}$.
The other (``constrained'') variables of~$J^{(r+1)}$ are expressed via unconstrained ones by means of using
the equations of~${\mathcal L}_{(r+1)}$.
We substitute the expressions for constrained variables into a conservation law and
split the obtained condition with respect to the unconstrained variables.
This procedure results in a first-order linear system of determining equations for conserved vectors.
Solving the determining equations up to the usual equivalence relation on $\CV(\mathcal{L})$,
we obtain complete description of local conservation laws of~$\mathcal{L}$.
To classify conservation laws in easier and more systematic way
(especially for classes of systems of differential equations), instead of usual equivalence
we use the introduced above equivalence with respect to symmetry or equivalence transformations.
(See Section~\ref{SectionLocalConsLwsOfDCEs} for examples.)

\medskip

\noindent {\em First iteration.}
After applying Lemma~\ref{lemma.null.divergence} to constructed conservation laws
on the set of solutions of~\mbox{$\mathcal{L}={\mathcal L}^0$},
we introduce potentials as additional dependent variables and attach the equations connecting the potentials with
components of corresponding conserved vectors to~${\mathcal L}^0$.
(If~\mbox{$n>2$} the attached equations of such kind form an underdetermined system with respect to the potentials.
Therefore, we can also attach gauge conditions on the potentials to~${\mathcal L}^0$.)

We have to use linear independent conservation laws since otherwise the introduced potentials will be
{\em dependent} in the following sense: there exists a linear combination of the potential tuples,
which is, for some $r'\in{\mathbb N}$, a tuple of functions of $x$ and $u_{(r')}$ only.

Then we exclude the unnecessary equations (i.e. the equations that are dependent on
equations of~${\mathcal L}^0$ and attached equations simultaneously)
from the extended (potential) system~${\mathcal L}^1$
which will be called a {\em potential system of the first level}.
Any conservation law of~${\mathcal L}^0$ is a one of~${\mathcal L}^1$.
We iterate the above procedure of the direct method for~${\mathcal L}^1$ to find its conservation laws
which are linear independent with ones from the previous iteration
and will be called {\em potential conservation laws of the first level}.

\medskip

\noindent {\em Further iterations.}
We make iterations while it is possible
(i.e. the iteration procedure has to be stopped if all the conservation laws of
a {\em potential system~${\mathcal L}^{k+1}$ of the $(k+1)$-th level} are linear dependent
with the ones of~${\mathcal L}^k$) or construct infinite chains of conservation laws by means of induction.
This process may yield {\em purely potential} conservation laws of the initial system~$\mathcal L$,
which are linear independent with local conservation laws and, therefore, depend explicitly on potential variables.

Any conservation law from the previous step of iteration procedure will be a conservation law for the next step
and vice versa, conservation laws which are obtained on the next step
and depend only on variables of the previous step are linear dependent with
conservation laws from the previous step.
It is also obvious that the conservation laws used for construction of a potential system of the next level are
trivial on the manifold of this system.

Since gauge conditions on potentials can be chosen in many different ways,
exhaustive realization of iterations is improbable in the case $n>2$.

\medskip

The procedure of exclusion of constrained variables (which are described above in detail only for the zeroth iteration)
is called in classical group analysis as ``confining to the mani\-fold of~$\mathcal L$''.
Taking into account~$\mathcal L$ in the above way, we automatically eliminate the ambiguity
connected with vanishing conserved vectors on the solutions of~$\mathcal L$.
However, the second kind ambiguity arising via existence of null divergences is preserved, and
it is the main reason of difficulties in realization of this algorithm with symbolic
computation systems~\cite{Wolf2002}.

To find conservation laws on each step of iteration procedure, one can apply other methods
which are based on the characteristic form~\eqref{CharFormOfConsLaw}
or its consequences~\eqref{NSCondOnChar} and~\eqref{NCondOnChar}.
These methods are also called as direct~\cite{Anco&Bluman2002a,Anco&Bluman2002b}.
Following~\cite{Wolf2002}, for convenience we will numerate them as
the second, third and fourth versions of the direct method in contrast to the above first one.
They are close to the symmetry group method by Noether since in the case of Euler--Lagrange equations
the coefficients~$\lambda^a$ are nothing else than Noether's characteristics.
Taking into account the equivalence relation on~$\Ch(\mathcal{L})$,
one can assume during calculations that characteristics depend only on unconstrained variables.

In the second version of the direct method
the representation~\eqref{CharFormOfConsLaw} is regarded as an equation defined on an open subset of~$J^{(k)}$
with respect to conserved vectors and characteristics simultaneously.

In the framework of the third version, sought quantities are characteristics only.
Determining equation~\eqref{NSCondOnChar} is defined on an open subset of~$J^{(k)}$.
Conserved vectors are reconstructed from known characteristics via explicit integral formulas.
An algorithm of this (third) version of the direct method was developed for Cauchy--Kovalevskaya systems
by S.~Anco and G.~Bluman~\cite{Anco&Bluman2002a,Anco&Bluman2002b}
(see also~\cite{Zharinov1986,Olver1986} for a theoretical background).

The fourth version is based on equation~\eqref{NCondOnChar} which is defined on the manifold~$\mathcal{L}$ and
is only a necessary condition on
characteristic of conservation laws. Therefore, one has to choose characteristics from the set of adjoint symmetries
using additional conditions.
Such approach was used by G.~Bluman and P.~Doran-Wu~\cite{Bluman&Doran-Wu1995}
for studying conservation laws of diffusion equations.

Each from four above versions of the direct method has its advantages and disadvantages.
A detailed comparative analysis of all the versions and
their realizations in computer algebra programs are given by T.~Wolf~\cite{Wolf2002}.

Prototypes and sources of a number of above ideas can be found in~\cite{Lax1968}.

\section{Two-dimensional case}\label{Section2DimCase}

The case of two independent variables is singular, in particular, with respect to possible (constant) indeterminacy
after introduction of potentials and high effectiveness of application of potential symmetries.
That is why we consider some notions connected with conservation laws in this case separately.
We denote independent variables as $t$ (the time variable) and $x$ (the space one).
Any local conservation law has the form
\begin{equation}\label{conslaw}
D_tF(t,x,u_{(r)})+D_xG(t,x,u_{(r)})=0.
\end{equation}
Here $D_t$ and $D_x$ are the operators of the total differentiation with respect to $t$ and $x$.
$F$ and $G$ are called the {\em conserved density} and the {\em flux} of the conservation law correspondingly.

Two conserved vectors $(F,G)$ and $(F',G')$ are {\em equivalent} if
there exist such functions~$\hat F$, $\hat G$ and~$H$ of~$t$, $x$ and derivatives of~$u$ that
$\hat F$ and $\hat G$ vanish on~$\mathcal{L}_{(k)}$ for some~$k$~and
\begin{equation}\label{2DConsLawEquivCondition}
F'=F+\hat F+D_xH ,\qquad G'=G+\hat G-D_tH.
\end{equation}

Any conservation law~\eqref{conslaw} of~$\mathcal{L}$ allows us to deduce the new dependent (potential) variable~$v$
by means of the equations
\begin{equation}\label{potsys1}
v_x=F,\qquad v_t=-G.
\end{equation}
To construct a number of potentials in one step,
we have to use a set of linear independent conservation laws (see the previous section)
since otherwise the potentials will be dependent in the following sense:
there exists a linear combination of the potentials,
which is, for some $r'\in{\mathbb N}$, a function of $t$, $x$ and $u_{(r')}$ only.

In the case of two independent variables we can also introduce the more general notion of
potential dependence.

\begin{definition}\label{DefinitionOfPotentialDependence}
The potentials $v^1$, \ldots, $v^p$ are called
{\em locally dependent on the set of solution of the system~${\mathcal L}$} (or, briefly speaking, {\em dependent})
if there exist $r'\in{\mathbb N}$ and a function~$H$ of the variables $t$, $x$, $u_{(r')}$, $v^1$, \ldots, $v^p$
such that $H(t,x,u_{(r')},v^1,\ldots,v^p)=0$ for any solution $(u,v^1,\ldots,v^p)$ of the united system determining
the set of potentials~$v^1$, \ldots, $v^p$.
\end{definition}

Proof of local dependence or independence of potentials for general classes of differential equations is difficult since
it is closely connected with precise description of possible structure of conservation laws.
An example of such proof for diffusion--convection equations is presented below.

In the case of single equation~$\mathcal{L}$, equations of form~\eqref{potsys1} combine into
the complete potential system since~$\mathcal{L}$ is a differential consequence of~\eqref{potsys1}.
As a rule, systems of such kind admit a number of nontrivial symmetries and so they are of a great interest.

Equations~\eqref{eq.tr.var.cons.law} and~\eqref{potsys1} imply the following statement.

\begin{proposition}\label{2DConsLawEquivRelation}
Any point transformation connecting two systems~$\mathcal{L}$ and~$\tilde{\mathcal L}$
of PDEs with two independent variables generates a one-to-one mapping between the sets of potential systems,
which correspond to~$\mathcal{L}$ and~$\tilde{\mathcal L}$. Generation is made via trivial prolongation
on the space of introduced potential variables, i.e. we can assume that the potentials are not transformed.
\end{proposition}

\begin{corollary}
The Lie symmetry group of a system~$\mathcal{L}$ of differential equations generates an equivalence group
on the set of potential systems corresponding to~$\mathcal{L}$.
\end{corollary}

\begin{corollary}
Let $\widehat{\mathcal{L}}|_S$ be the set of all potential systems constructed
for systems from the class~$\mathcal{L}|_S$ with their conservation laws.
Action of transformations from~$G^{\Equiv}(L,S)$ together with the equivalence relation of potentials
naturally generates an equivalence relation on~$\widehat{\mathcal{L}}|_S$.
\end{corollary}

\begin{note}
Proposition~\ref{2DConsLawEquivRelation} and its Corollaries imply that the equivalence group for a class of
systems or the symmetry group for single system can be prolonged to potential variables for any step of
the direct iteration procedure. It is natural the prolonged equivalence groups are used to classify
possible conservation laws and potential systems in each iteration.
Additional equivalences which exist in some subclasses of the class or arise
after introducing potential variables can be used for deeper analysis of connections between conservation laws.
\end{note}

\section{Local conservation laws of diffusion--convection equations}\label{SectionLocalConsLwsOfDCEs}

To classify the conservation laws of equations from class~\eqref{eqf1}
we have to start our investigation from finding equivalence transformations.
Application of the direct method to class~(\ref{eqf1}) allows us to
find the complete equivalence group $G^{\Equiv}$
including the both continuous and discrete transformations.
The following statement is true.

\begin{theorem}[see~\cite{Popovych&Ivanova2004NVCDCEs,Popovych&Ivanova2005PETs}]
\label{TheoremNDCEsGequiv}
Any transformation from $G^{\Equiv}$ has the form
\[
\tilde t=\varepsilon_4t+\varepsilon_1, \quad
\tilde x=\varepsilon_5x+\varepsilon_7 t+\varepsilon_2, \quad
\tilde u=\varepsilon_6u+\varepsilon_3, \quad
\tilde A=\varepsilon_4^{-1}\varepsilon_5^2A, \quad
\tilde B=\varepsilon_4^{-1}\varepsilon_5B-\varepsilon_7,
\]
where $\varepsilon_1,$ \dots, $\varepsilon_7$ are arbitrary constants,
$\varepsilon_4\varepsilon_5\varepsilon_6\ne0.$
\end{theorem}

The kernel of Lie symmetry group of equations from class~\eqref{eqf1} is
the group of the transformations which are common for all of these equations.
We denote it as $G^{\ker}$. Via trivial prolongation on the arbitrary elements~$A$ and~$B$,
$G^{\ker}$ is isomorphic to a normal subgroup of~$G^{\Equiv}$.

\begin{theorem}\label{TheoremNDCEsGker}
$G^{\ker}$ is formed by the transformations
$\tilde t=t+\varepsilon_1$,
$\tilde x=x+\varepsilon_2$,
$\tilde u=u$,
where $\varepsilon_1$ and $\varepsilon_2$ are arbitrary constants.
\end{theorem}

First we search the conservation laws of equations from class~(\ref{eqf1})
in the form~\eqref{conslaw}.

\begin{lemma}\label{LemmaOnOrderOfConsLawsOfDCEs}
Any local conservation law of any equation from class~\eqref{eqf1} has the first order.
Moreover, up to equivalence on conserved vectors one can assume that
the density depending on $t$, $x$ and $u$ and the flux depending on $t$, $x$, $u$ and $u_x$.
\end{lemma}

\begin{proof}
Considering conservation laws on the manifold of equation~\eqref{eqf1} and
its differential consequences, we can assume that $F$ and $G$ depend only on $t$, $x$ and
$u_k=\partial^k u/\partial x^k$, $k=\overline{0,r'},$ where $r'\le 2r$.
We expand the total derivatives in (\ref{conslaw}) and take into account differential consequences of the
form $u_{tj}=D_x^{j+2}\int\! A+D_x^{j+1}\int\! B$, where $\int\! A=\int\! A(u) du,$ $\int\! B=\int\! B(u)du,$
$j=\overline{0,r'}$. As a result we obtain the following condition
\begin{equation}\label{clcdeom}\textstyle
F_t+F_{u_j}(D_x^{j+2}\int\! A+D_x^{j+1}\int\! B)+G_x+G_{u_j}u_{j+1}=0.
\end{equation}
Let us decompose~(\ref{clcdeom}) with respect to the highest derivatives $u_j$.
So, the coefficients of $u_{r'+2}$ and $u_{r'+1}$ give the equations
$F_{u_{r'}}=0$, $G_{u_{r'}}+AF_{u_{r'-1}}=0$ that result in
\[
F=\hat F, \quad G=-Au_{r'}\hat F_{u_{r'-1}}+\hat G,
\]
where $\hat F$ and $\hat G$ are functions of $t$, $x$, $u$, $u_1$, \ldots, $u_{r'-1}$.
Then, after selecting the terms containing $u_{r'}^2$, we obtain that $-A\hat F_{u_{r'-1}u_{r'-1}}=0$.
It yields that $\hat F =\check F^1u_{r'-1}+\check F^0,$
where $\check F^1$ and $\check F^0$ depend only on $t$, $x$, $u$, $u_1$,~\ldots, $u_{r'-2}$.

Consider the conserved vector with the density~$\tilde F=F-D_xH$ and the flux~$\tilde G=G+D_tH$,
where $H=\int \check F^1du_{r'-2}$. This conserved vector is equivalent to the initial one, and
\[
\tilde F=\tilde F(t,x,u,u_1,\ldots,u_{r'-2}), \quad
\tilde G=\tilde G(t,x,u,u_1,\ldots,u_{r'-1}).
\]
Iterating this procedure a necessary number of times, we obtain the lemma statement.
\end{proof}

\begin{note}
A similar statement is true for an arbitrary (1+1)-dimensional evolution equation~$\cal L$ of the even
order~$r=2\bar r$, $\bar r\in\mathbb{N}$~\cite{Abellanas&Galindo1979,Ibragimov1985}.
For example~\cite{Ibragimov1985}, for any conservation law of~$\cal L$
we can assume up to equivalence of conserved vectors
that $F$ and $G$ depend only on~$t$, $x$ and derivatives of~$u$ with respect to~$x$, and
the maximal order of derivatives in~$F$ is not greater than $\bar r$.

Lemma~\ref{LemmaOnOrderOfConsLawsOfDCEs} gives a stronger result for a more restricted class of equations.
In the above proof we specially use the most direct method to demonstrate its effectiveness in
quite general cases. This proof can be easily extended to other classes of (1+1)-dimensional evolution equations
of odd orders and potential systems corresponding to equations from class~\eqref{eqf1}
(see the proof of Lemma~\ref{LemmaRedConsLaws}).
\end{note}

\begin{theorem}\label{TheoremCLCLofDCEs}
Any equation from class~\eqref{eqf1} has the conservation law~\eqref{conslaw} where
\begin{equation}\textstyle\label{ker.cons.law}
1.\quad F=u,\qquad G=-Au_x-\int\! B.
\end{equation}
A complete list of $G^{\Equiv}$-inequivalent equations~\eqref{eqf1} having
additional (i.e. linear independent with~\eqref{ker.cons.law}) conservation laws
is exhausted by the following ones
\begin{gather}\textstyle
2.\quad \forall A, \quad B=0: \qquad F=xu, \quad G=\int\! A-xAu_x,  \label{conslawB0}\\\textstyle
3.\quad \forall A, \quad B=A:
 \qquad F=(e^x+\varepsilon)u, \quad G=-(e^x+\varepsilon)Au_x-\varepsilon \int\! A,  \label{conslawBA}\\\textstyle
4.\quad A=1, \quad B=0: \qquad F=\alpha u, \quad G=\alpha_xu-\alpha u_x,  \label{conslawA1B0}
\end{gather}
where $\varepsilon\in\{0,\pm1\}\!\!\mod G^{\Equiv}$, $\int\! A=\int A(u) du$, $\int\! B=\int\! B(u) du$,
$\alpha=\alpha(t,x)$ is an arbitrary solution of the linear heat equation $\alpha_t+\alpha_{xx}=0$.
(Together with values $A$ and $B$ we also adduce complete lists of
densities and the fluxes of additional conservation laws.)
\end{theorem}

\begin{proof}
In view of Lemma~\ref{LemmaOnOrderOfConsLawsOfDCEs}, we can assume at once that $F=F(t,x,u)$ and $G=G(t,x,u,u_x)$.
Let us substitute the expression for~$u_t$ deduced from~(\ref{eqf1})
into~(\ref{conslaw}) and decompose the obtained equation with respect to $u_{xx}$. The coefficient of~$u_{xx}$
gives the equation $AF_u+G_{u_x}=0$, therefore $G=-AF_{u}u_x+G^1(t,x,u)$.
Taking into account the latter expression for $G$ and splitting the rest of equation~(\ref{conslaw})
with respect to the powers of $u_x$, we obtain the system of PDEs on the functions
$F$ and $G^1$ of the form
\begin{equation}\label{splitconslaw}
F_{uu}=0, \quad BF_u-AF_{ux}+G^1_u=0, \quad F_t+G^1_x=0.
\end{equation}
Solving first two equations of~(\ref{splitconslaw}) yields
\begin{equation}\label{efFaG}\textstyle
F=F^1(t,x)u+F^0(t,x), \quad G^1=F^1_x\int\! A-F^1\int\! B+G^0(t,x).
\end{equation}
In further consideration the major role is played by the equation~$AF_{uxx}-BF_{ux}+F_{ut}=0$
that is a differential consequence of system~(\ref{splitconslaw}) and can be rewritten as
\[
AF^1_{xx}-BF^1_x+F^1_t=0.
\]
Indeed, it is the unique classifying condition for this problem.
There exist four different possibilities for values $A$ and $B$.
In all cases we obtain the equation $F^0_t+G^0_x=0$.
Therefore, up to conservation laws equivalence we can assume $F^0=G^0=0$.
Moreover, the function $F^1={\rm const}$ is a solution of the classifying condition in the general case.
This solution corresponds to the conservation laws of Case~1.
Only conservation laws of such kind exist for all admissible values of arbitrary elements~$A$ and~$B$.
Then we classify special values of $A$ and $B$ for which equation~(\ref{eqf1}) possesses additional
conservation laws.

\vspace{0.8ex}

1. $B\not\in\langle A,1\rangle$. Then $F^1_x=0$, $F^1_t=0$
that gives contradiction with the assumption $F^1\ne{\rm const}$.

\vspace{0.8ex}

2. $A\not\in\langle 1\rangle$, $B\in\langle 1\rangle$.
Then $B=0\!\mod G^{\Equiv}$ and $F^1_{xx}=0,$ $F^1_t=0$, i.e. $F^1=x\!\mod G^{\Equiv}$ (Case~2).

\vspace{0.8ex}

3. $B\in\langle A,1\rangle$, $A,B\not\in\langle 1\rangle$.
Then $B=A\!\mod G^{\Equiv}$ and $F^1_{xx}+F^1_x=0$, $F^1_t=0$,
i.e. $F^1=e^x+\varepsilon\!\mod G^{\Equiv}$, where $\varepsilon\in\{0,\pm 1\}$ (Case~3).

\vspace{0.8ex}

4. $A,B\in\langle 1\rangle$. Therefore, $A=1$, $B=0\!\mod G^{\Equiv}$ and $F^1_t+F^1_{xx}=0$ (Case~4).
\end{proof}

\begin{note}
It follows from the proof that we can assume $(A,B)\not={\rm const}$ in Cases~1, 2 and~3.
(If $(A,B)={\rm const}$ Cases~1, 2 and~3 are included in Case~4 for different values of~$\alpha$.)
\end{note}

Using the conservation laws adduced in Theorem~\ref{TheoremCLCLofDCEs}, we can introduce potentials for
different values of the parameter-functions~$A$ and~$B$ and construct the corresponding potential systems
(Cases~1--4 of Table~1). The important question for our consideration is whether the introduced potentials
are locally independent in the sense of Definition~\ref{DefinitionOfPotentialDependence}.
If we know the precise structure of conservation laws the answer is almost obvious.

\begin{theorem}\label{TheoremOnIndepPotOfEvolEqs}
For any equation~\eqref{eqf1} potentials are locally dependent on the equation manifold
iff the corresponding conservation laws are linear dependent.
\end{theorem}

\begin{proof}
Since the direct statement of the theorem is obvious (see Section~\ref{Section2DimCase}),
we prove only the inverse statement, using the rule of contraries.
Suppose that potentials~$v^0$, \ldots, $v^p$ introduced with (independent) conservation laws of Cases~1--4
are locally dependent. The vanishing $p$ means local triviality of $v^0$ as a potential,
i.e. $v^0$ can be expressed in terms of local variables and the corresponding conservation law is trivial.
That is why it is sufficient to investigate only the special cases when the number of independent conservation laws
is greater than~1. Therefore, $p=1$ if either $B=0$ or $B=A$ and $p\in\mathbb N/\{0\}$ for the linear heat equation.

Without loss of generality we can assume that
there exist $r\in\mathbb N$ and a fixed function $V$ of~$t$, $x$, $\bar v =(v^1,\ldots v^p)$ and $u_{(r)}$ that
$v^0=V(t,x,\bar v,u_{(r)})$ for any solution of the united system determining
the whole set of potentials~$v^0$, \ldots, $v^p$.
Taking into account equation~\eqref{eqf1} and its differential consequences,
we can assume that $V$ depends only on $t$, $x$, $\bar v$ and
$u_k=\partial^k u/\partial x^k$, $k=\overline{0,r'},$ where $r'\le 2r$.
Let us apply the operator~$D_x$ to the condition $v^0=V(t,x,\bar v,u,u_1,\ldots,u_{r'})$:
$v^0_x=V_x+V_{v^s}v^s_x+V_{u^k}u_{k+1}$. (Hereafter the index~$s$ takes the values from~1 to~$p$.)
Since in any case under consideration $v^s_x=f^su$ for some functions $f^s$ of $t$ and $x$,
we can split the differentiated condition with respect to $u_k$ step-by-step in the reverse sequence,
beginning with the major derivative. As a result, we obtain $V_{u^k}=0$, $V_x=0$ and $f^0=V_{v^s}f^s$,
i.e. the functions~$f^0$, \ldots $f^p$ are linear dependent.
This gives a contradiction with the supposition that the conservation laws are independent.
\end{proof}

\section{Simplest potential conservation laws \\ of diffusion--convection equations}
\label{SectionSimplestPotentialConservationLaws}

Let us investigate local conservation laws of potential systems~1--4 from Table~1, which have the form\vspace{-1.5ex}
\begin{equation}\label{conslaw2}
D_tF(t,x,u_{(r)},v_{(r)})+D_xG(t,x,u_{(r)},v_{(r)})=0.
\end{equation}
These laws can be considered as nonlocal ({\em potential}) conservation laws of equations from class~(\ref{eqf1}).
We assume them as simplest potential conservation laws since the corresponding potential systems are constructed
from 
the initial equation~(\ref{eqf1}) with one conservation law only.

We classify conservation laws up to the equivalence relation with respect to
the transformation group~$G^{\Equiv}_{\rm pr}$ which is a result of the trivial prolongation of the group~$G^{\Equiv}$
from Theorem~\ref{TheoremNDCEsGequiv} to the space of the potential~$v$.

\begin{lemma}\label{LemmaRedConsLaws}
Any conservation law of form~\eqref{conslaw2}
for each of systems~1--4 from Table~1 has the zero order,
i.e. it is equivalent to a law with a conserved density $F$ and a flux $G$
that are independent on the (non-zero order) derivatives of $u$ and $v$.
\end{lemma}

\begin{proof}
Consider any from the systems~1--4.
Taking it and its differential consequences into account, we can exclude
dependence of $F$ and $G$ on the all (non-zero order) derivatives of $v$ and the derivatives of $u$
containing differentiation with respect to $t$.
The remain part of the proof is completely similar to the one of Lemma~\ref{LemmaOnOrderOfConsLawsOfDCEs}.
\end{proof}

In an analogous to Theorem~\ref{TheoremCLCLofDCEs} and more cumbersome way
we prove the following statement.

\begin{theorem}\label{theorconslaw2}
A complete set of $G^{\Equiv}_{\rm pr}$-inequivalent conservation laws
of form~\eqref{conslaw2} for equations from class~\eqref{eqf1}
is presented in Table~1 with a~double numeration of cases.
\end{theorem}

\newcounter{tbnIvanova} \setcounter{tbnIvanova}{0}
\begin{center}\small
{\bf Table~1.} Conservation laws and potential systems of convection-diffusion equations. \\[1ex]
\footnotesize\renewcommand{\arraystretch}{1.2}
\begin{tabular}{|l|c|c|c|c|l|}
\hline\raisebox{0ex}[2.8ex][0ex]{\null}
N & $A$ &$B$ &  $F$& $G$ & \hfill {Potential system \hfill} \\
\hline\raisebox{0ex}[2.8ex][0ex]{\null}
\refstepcounter{tbnIvanova}\thetbnIvanova&$\forall$ & $\forall$ & $u$ & $-Au_x-\int\! B$ & $v_x=u,\ v_t=Au_x+\int\! B$
\raisebox{0ex}[0ex][1.2ex]{\null}\\[-2.3ex]\multicolumn{6}{|c|}%
{\hspace*{-1.8ex}\dotfill\hspace*{-2ex}}\\[-1.05ex]\raisebox{0ex}[2.8ex][0ex]{\null}\thetbnIvanova.1
&$\forall$ & $0$ & $v$ & $-\int\! A$ & $v_x=u,\ w_x=v,\ w_t=\int\! A$
\\
\thetbnIvanova.2&
$\forall$ & $A$ & $e^xv$&$-e^x\int\! A$
      & $v_x=u,\ w_x=e^xv,\ w_t=e^x\int\! A$
\\
\thetbnIvanova.3&$\forall$ & $\int\! A+uA$ & $e^v$ & $-e^v\int\! A$ & $v_x=u,
         \ w_x=e^v,\ w_t=e^v\int\! A$
\\
\thetbnIvanova.4&$u^{-2}$ & $0$ & $\sigma$ & $\sigma_vu^{-1}$ & $v_x=u,
\ w_x=\sigma,\ w_t=-\sigma_vu^{-1}$
\\
\thetbnIvanova.5&$u^{-2}$ & $u^{-2}$ & $\sigma e^x$ & $\sigma_vu^{-1}e^x$ & $v_x=u,
    \ w_x=\sigma e^x,\ w_t=-\sigma_vu^{-1}e^x$
\\
\thetbnIvanova.6&$1$ & $2u$ & $\alpha e^v$ & $\alpha_xe^v-\alpha u e^v$ & $v_x=u,
    \ w_x=\alpha e^v,\ w_t=\alpha u e^v-\alpha_xe^v$
\\[0.3ex] \hline\raisebox{0ex}[2.8ex][0ex]{\null}
\refstepcounter{tbnIvanova}\thetbnIvanova&$\forall$ & $0$ & $xu$ & $\int\! A-xAu_x$ & $v_x=xu,\ v_t=xAu_x-\int\! A$
\raisebox{0ex}[0ex][1.2ex]{\null}\\[-2.3ex]\multicolumn{6}{|c|}%
{\hspace*{-1.8ex}\dotfill\hspace*{-2ex}}\\[-1.05ex]\raisebox{0ex}[2.8ex][0ex]{\null}\thetbnIvanova.1&
$\forall$ & $0$ & $x^{-2}v$ & $-x^{-1}\int\! A$ &
     $v_x=xu,\ w_x=x^{-2}v,\ w_t=x^{-1}\int\! A$
\\[0.3ex] \hline\raisebox{0ex}[2.8ex][0ex]{\null}
\refstepcounter{tbnIvanova}\thetbnIvanova&$\forall$ & $A$ & $(e^x+\varepsilon)u$ & $-(e^x+\varepsilon)Au_x-
\varepsilon\int\! A$ &
     $v_x=(e^x+\varepsilon)u,\ v_t=(e^x+\varepsilon)Au_x+\varepsilon\int\! A\!\!$
\raisebox{0ex}[0ex][1.2ex]{\null}\\[-2.3ex]\multicolumn{6}{|c|}{\hspace*{-1.8ex}\dotfill\hspace*{-2ex}}\\[-1.05ex]
\raisebox{0ex}[6.8ex][0ex]{\null}\thetbnIvanova.1 
&$\forall$ & $A$ & $\dfrac{e^x}{(e^x+\varepsilon)^2}v$
       & $-\dfrac{e^x}{e^x+\varepsilon}\int\!A$ &
      $\renewcommand{\arraycolsep}{0ex}\begin{array}{l}v_x=(e^x+\varepsilon)u,\ w_x=\dfrac{e^x}{(e^x+\varepsilon)^2}v,\\
       w_t=\dfrac{e^x}{e^x+\varepsilon}\int\!A  \end{array} $
\\[4.3ex] \hline\raisebox{0ex}[2.8ex][0ex]{\null}
\refstepcounter{tbnIvanova}\thetbnIvanova&$1$ & $0$ & $\alpha u$ &
$\alpha_xu-\alpha u_x$ & $v_x=\alpha u,\ v_t=\alpha u_x-\alpha_xu$
\raisebox{0ex}[0ex][1.2ex]{\null}\\[-2.3ex]\multicolumn{6}{|c|}%
{\hspace*{-1.8ex}\dotfill\hspace*{-2ex}}\\[-1.05ex]\raisebox{0ex}[5.6ex][0ex]{\null}\thetbnIvanova.1
& $1$ & $0$ & $\left(\dfrac{\beta}{\alpha}\right)_x v$
  & $-\alpha\left(\dfrac{\beta}{\alpha}\right)_x\! u-\left(\dfrac{\beta}{\alpha}\right)_tv$
   & $\renewcommand{\arraycolsep}{0ex}\begin{array}{l}v_x=\alpha u,
      \\[1ex]w_x=\left(\dfrac{\beta}{\alpha}\right)_x\! v,\quad
      w_t=\alpha\left(\dfrac{\beta}{\alpha}\right)_x\! u+\left(\dfrac{\beta}{\alpha}\right)_t\! v\end{array}$
\\[3.6ex]
\hline
\end{tabular}\\[2.5ex]
\parbox{160mm}{\footnotesize
Here $\varepsilon\in\{0,\pm 1\},$ $\int\! A=\int\! A(u)du$, $\int\! B=\int\! B(u)du$;
$\alpha(t,x)$, $\beta(t,x)$ and $\sigma(t,v)$ are arbitrary solutions of the
linear heat equation ($\alpha_t+\alpha_{xx}=0$, $\beta_t+\beta_{xx}=0$,
$\sigma_t+\sigma_{vv}=0$).
In Case 1.3 we assume $\int B=u\int A$ for a conservation law to have the adduced form.\par
}
\end{center}

\begin{note}
To prove Theorem~\ref{theorconslaw2} we use all independent differential consequences
of correspondent potential systems. In Table~1 for the double numerated potential systems we omit equations
containing $v_t$ since they are only differential consequences of equations of these systems.
\end{note}

Let us analyze connections between conservation laws and ones between potential systems,
which arise due to additional (including purely potential) equivalence transformations in some special cases.
Below we assume $A\not\in\{1,u^{-2}\}\!\!\mod G^{\Equiv}$ as a general value of $A$.

\medskip

{\noindent\bf General case.} Equation~\eqref{eqf1} in the general case has
the unique linear independent local conservation law (Case~1) with the conserved vectors $(F^1,G^1)=(u,-Au_x)$.
All conservation laws of the corresponding potential system
\begin{equation} \label{PotSysGen}\textstyle
v^1_x=u,\quad v^1_t=Au_x+\int\! B,
\end{equation}
are trivial, i.e. in our framework equation~\eqref{eqf1} of the general form admits only trivial
potential conservation laws.

\medskip

{\noindent\mathversion{bold}$B=0$.} Any equation of such form has two linear independent local conservation laws
(Cases~1 and~2) with the conserved vectors $(F^1,G^1)=(u,-Au_x)$ and $(F^2,G^2)=(xu,\int\! A-xAu_x)$,
and any conservation law is $G^{\ker}$-equivalent to one of them.
Using these conservation laws, we can introduce two potentials $v^1$ and $v^2$, where
\begin{gather}
v^1_x=u,\quad v^1_t=Au_x, \label{potsysB0gen}\\[1ex]
\textstyle  v^2_x=xu,\quad v^2_t=xAu_x-\int\!A.  \label{potsysB0spec}
\end{gather}
Equations~\eqref{potsysB0gen} and~\eqref{potsysB0spec} considered separately form two potential systems
for equation~\eqref{eqf1} with vanishing $B$ in unknown functions $(u,v^1)$ and $(u,v^2)$ correspondingly.
The third potential system is formed by equations~\eqref{potsysB0gen} and~\eqref{potsysB0spec}
simultaneously, and three functions $u$, $v^1$ and $v^2$ are assumed as unknown.
Each from systems~\eqref{potsysB0gen} and~\eqref{potsysB0spec} has one linear independent local conservation law
(Cases~1.1 and~2.1).
These conservation laws with conserved vectors
$(F^{11},G^{11})=(v^1,-\int\!A)$ and $(F^{21},G^{21})=(x^{-2}v^2,-x^{-1}\int\!A)$
are simplest potential conservation laws for equation~\eqref{eqf1} with vanishing $B$
and allow us to introduce ``second-level'' potentials $w^1$ and $w^2$. As a result, we obtain two
potential systems of the next level:
\begin{gather}\textstyle
v^1_x=u,\quad w^1_x=v^1,\quad w^1_t=\int\!A,\label{potsys2B0gen}\\[1ex] \textstyle
v^2_x=xu,\quad w^2_x=x^{-2}v^2,\quad w^2_t=x^{-1}\int\!A.  \label{potsys2B0spec}
\end{gather}

At the same time, the simplest potential conservation laws are trivial on the solution manifold of
the united system~\eqref{potsysB0gen}--\eqref{potsysB0spec} since
\begin{gather*}
F^{11}=D_x(xv^1-v^2),\quad G^{11}=-D_t(xv^1-v^2),\\[1ex]
F^{21}=D_x(v^1-x^{-1}v^2),\quad G^{21}=-D_t(v^1-x^{-1}v^2).
\end{gather*}
Moreover $w^1=xv^1-v^2$, $w^2=v^1-x^{-1}v^2$,
i.e. systems~\eqref{potsys2B0gen}, \eqref{potsys2B0spec} and~\eqref{potsysB0gen}--\eqref{potsysB0spec}
are locally equivalent.
It implies that really system~\eqref{potsysB0gen}--\eqref{potsysB0spec}
is generated by only three independent equations.
We can choose e.g. the equations \[\textstyle v^1_x=u,\quad v^2_x=xu,\quad xv^1_t-v^2_t=\int\!A.\]
Although system~\eqref{potsys2B0gen} formally belongs to the second level,
it is the most convenient for further investigation since it has the simplest form.

\pagebreak

{\noindent\mathversion{bold}$B=A$.}
This case is analyzed in the similar way to the previous one.
Any equation with $B=A$ has the two-dimensional space of local conservation laws.
Up to $G^{\ker}$-equivalence, we have two possibilities for conserved vectors (Cases~1 and~3):
\begin{gather*}
\textstyle
(F^1,G^1)=(u,-Au_x-\int\! A)\quad \mbox{and}\quad\\[1.5ex]\textstyle
(F^3,G^3)=((e^x+\varepsilon)u,-(e^x+\varepsilon)Au_x-\varepsilon \int\!A).
\end{gather*}
Using these conservation laws, we can introduce two potentials $v^1$ and $v^3$, where
\begin{gather}\textstyle
v^1_x=u,\quad v^1_t=Au_x+\int\!A, \label{potsysBAgen}\\[1.5ex] \textstyle
v^3_x=(e^x+\varepsilon)u,\quad v^3_t=(e^x+\varepsilon)Au_x+\varepsilon \int\!A.  \label{potsysBAspec}
\end{gather}
Equations~\eqref{potsysBAgen} and~\eqref{potsysBAspec} considered separately form two potential systems
for equation~\eqref{eqf1} with $B=A$ in unknown functions $(u,v^1)$ and $(u,v^3)$ correspondingly.
The third potential system is formed by equations~\eqref{potsysBAgen} and~\eqref{potsysBAspec}
simultaneously, and three functions $u$, $v^1$ and $v^3$ are assumed as unknown.
Each from systems~\eqref{potsysBAgen} and~\eqref{potsysBAspec} has one linear independent local conservation law
(Cases~1.2 and~3.1).
These conservation laws with conserved vectors
\begin{gather*}\textstyle
(F^{12},G^{12})=(e^xv^1,-e^x\int\!A) \quad \mbox{and}\\[1.5ex] \textstyle
(F^{31},G^{31})=\left(\dfrac{e^x}{(e^x+\varepsilon)^2}v^3,-\dfrac{e^x}{e^x+\varepsilon}\int\!A\right)
\end{gather*}
are simplest potential conservation laws for equation~\eqref{eqf1} with $B=A$
and allow us to introduce ``second-level'' potentials $w^1$ and $w^3$. As a result, we obtain two
potential systems of the next level:
\begin{gather}\textstyle
v^1_x=u,\quad w^1_x=e^xv^1,\quad w^1_t=e^x\int\!A,\label{potsys2BAgen}\\[1ex] \textstyle
v^3_x=(e^x+\varepsilon)u,\quad w^3_x=\dfrac{e^x}{(e^x+\varepsilon)^2}v^3,
\quad w^3_t=\dfrac{e^x}{e^x+\varepsilon}\int\!A.  \label{potsys2BAspec}
\end{gather}

At the same time, the simplest potential conservation laws are trivial on the solution manifold of
the united system~\eqref{potsysBAgen}--\eqref{potsysBAspec} since
\begin{gather*}
F^{12}=D_x((e^x+\varepsilon)v^1-v^3), \quad G^{12}=-D_t((e^x+\varepsilon)v^1-v^3),\\[1.5ex]
F^{31}=D_x\left(v^1-\dfrac{v^3}{e^x+\varepsilon}\right), \quad G^{31}=-D_t\left(v^1-\dfrac{v^3}{e^x+\varepsilon}\right).
\end{gather*}
Moreover
\[
w^1=(e^x+\varepsilon)v^1-v^3,\quad w^3=v^1-\dfrac{v^3}{e^x+\varepsilon},
\]
i.e. systems~\eqref{potsys2BAgen}, \eqref{potsys2BAspec} and~\eqref{potsysBAgen}--\eqref{potsysBAspec}
are locally equivalent.
It implies that really system~\eqref{potsysBAgen}--\eqref{potsysBAspec}
is generated by only three independent equations.
We can choose e.g. the equations
\[\textstyle
v^1_x=u,\quad v^3_x=(e^x+\varepsilon)u,\quad (e^x+\varepsilon)v^1_t-v^3_t=e^x\int\!A.
\]
Although system~\eqref{potsys2BAgen} formally belongs to the second level,
it is the most convenient for further investigation since it has the simplest form.

\medskip

{\noindent\mathversion{bold}$B=\int\! A+uA$.} The initial potential system in Case~1.3 is reduced to Case~1.2
by means of the hodograph transformation
\begin{equation}\label{HodographXV}
 \tilde t=t, \quad \tilde x=v, \quad \tilde v=x, \quad \tilde u=u^{-1}, \quad \tilde A=u^{-2}A,
\end{equation}
and the conservation law~1.3 is transformed to the local one of Case~3 where $\varepsilon=0$.
The same transformation extended to the potential~$w$ as $\tilde w=-w+ve^x$ also reduces
the potential system~1.3 to~1.2.

\medskip

{\noindent\bf Linearizable equations.}
It is well known~\cite{Bluman&Kumei1980,Storm1951,Forsyth1906,Hopf1950,Cole1951,Fokas&Yortsos1982,Strampp1982b}
that equations~\eqref{eqf1} $G^{\Equiv}$-equi\-valent to ones of Cases~1.4, 1.5, 1.6 are linearized by a nonlocal
(so-called potential equivalence \cite{Popovych&Ivanova2005PETs,LisleDissertation})
transformations to the linear heat equation.
That is why these equations stand out against the other diffusion--convection equations
with having an infinite number of linear independent purely potential conservation laws.

The $u^{-2}$-diffusion equation $u_t=(u^{-2}u_x)_x$
is reduced to the linear heat equation~\cite{Bluman&Kumei1980}
by the hodograph transformation~\eqref{HodographXV}.
More exactly, \eqref{HodographXV} is a local transformation between
the corresponding potential systems $v_x=u$, $v_t=u^{-2}u_x$ and $v_x=u$, $v_t=u_x$
constructed by means of using the ``common'' conservation law (Case~1).
The $u^{-2}$-diffusion equation has, as a subcase of the case~$B=0$,
two linear independent local conservation laws with the conserved vectors
\begin{gather}
F^1=u,\quad G^1=-u^{-2}u_x,\label{commonclu-2e}\\[1ex]
F^2=xu,\quad G^2=-xu^{-2}u_x-u^{-1}\label{specclu-2e}
\end{gather}
(Cases~1 and~2 of Table~1)
and the infinite series of potential conservation laws (Case~1.4) additionally.
Under the action of hodograph transformation~\eqref{HodographXV}
the conservation law with conserved vector~\eqref{commonclu-2e}
is transformed to the trivial one of the linear heat equation
with the conserved vector~$(1,0)$. And~vice versa, the conservation law of the linear heat equation corresponding to
the value $\alpha=1$ (Case~4) is transformed by~\eqref{HodographXV} to the trivial one of the $u^{-2}$-diffusion equation
with the conserved vector~$(1,0)$. The conservation law with conserved vector~\eqref{specclu-2e} is trivial
on the manifold of potential system constructed by means of~\eqref{commonclu-2e}, is equivalent to the one from
Case~1.4 with $\sigma=1$ and is transformed to Case~4.1, where $\alpha=1$ and $\beta=x$.
Case~1.4 is reduced by~\eqref{HodographXV} to Case~4, where $\alpha=\sigma$.

Since the equation $u_t=(u^{-2}u_x)_x+u^{-2}u_x$ is reduced to the $u^{-2}$-diffusion equation
by the local transformation $\tilde t=t$, $\tilde x=e^x$, $\tilde u=e^{-x}u$,
its conservation laws are connected with ones of the linear heat equation in
a way which is similar to the previous case.

The potential systems $\tilde v_{\tilde x}=\tilde u$, $\tilde v_{\tilde t}=\tilde u_{\tilde x}+\tilde u^2$
and $v_x=u$, $v_t=u_x$ constructed with the ``common'' conservation law
for the Burgers equation~$u_t=u_{xx}+2uu_x$ and the linear heat equation~$u_t=u_{xx}$
are connected via the transformation
\[
t=\tilde t,\quad x=\tilde x,\quad u=\tilde u e^{\tilde v},\quad v=e^{\tilde v}.
\]
(Here the tilde variables correspond to the Burgers equation.)
Let us note that really the famous Cole--Hopf transformation~\cite{Cole1951,Hopf1950}
(first found in~\cite{Forsyth1906}) linearizes the Burgers equation
to the ``potential'' heat equation $v_t=v_{xx}$~\cite{LisleDissertation,Popovych&Ivanova2005PETs}.
The Burgers equation has the ``common'' local conservation law (Case~1)
and the infinite series of simplest potential conservation laws (Case~1.6).
The above transformation between the potential systems induces the one-to-one mapping preserving~$\alpha$
between the infinite series~1.6 and the one~4 of the ``potential'' heat equation $v_t=v_{xx}$.
Then, the conservation law of form 4 with the function~$\tilde\alpha$ for the ``potential'' heat equation $v_t=v_{xx}$
is equivalent to the one with the function~$\alpha$ for the heat equation $u_t=u_{xx}$,
where $\tilde\alpha=\alpha_x$.
The conservation law of Case~1 for the Burgers equation is trivial on the manifold of
the corresponding potential system and is mapped to trivial one of the system $v_x=u$, $v_t=u_x$.

\pagebreak

{\noindent\bf Linear heat equation.} The linear heat equation $u_t=u_{xx}$
has an infinite dimensional space of local conservation laws~\cite{Dorodnitsyn&Svirshchevskii1983},
which is generated by conserved vectors of the form \[(F^{\alpha},G^{\alpha})=(\alpha u,\alpha_xu-\alpha u_x),\]
where $\alpha=\alpha(t,x)$ is an arbitrary solution of the backward linear heat equation $\alpha_t+\alpha_{xx}=0$.
Using a fixed conservation law of such form, we can introduce the potential $v^{\alpha}$, where
\begin{equation}\label{potsysLin}
v^{\alpha}_x=\alpha u,\quad v^{\alpha}_t=\alpha u_x-\alpha_xu.
\end{equation}
System~\eqref{potsysLin} has one infinite series of conservation laws (Case~4.1) with conserved vectors
\begin{equation}\label{dvcllheAB}
(F^{\alpha\beta},G^{\alpha\beta})=\left(\left(\dfrac{\beta}{\alpha}\right)_x v^{\alpha},
 -\alpha\left(\dfrac{\beta}{\alpha}\right)_x\! u-\left(\dfrac{\beta}{\alpha}\right)_tv^{\alpha}\right),
\end{equation}
where $\beta=\beta(t,x)$ is an arbitrary solution of the backward linear heat equation $\beta_t+\beta_{xx}=0$.
These conservation laws are simplest potential ones for the linear heat equation
and allow us to introduce ``second-level'' potentials $w^{\alpha\beta}$. As a result, we obtain
potential systems of the next level:
\begin{equation}\label{potsysLinsec}
v^{\alpha}_x=\alpha u, \qquad
w^{\alpha\beta}_x=\left(\dfrac{\beta}{\alpha}\right)_x v^{\alpha},\qquad
w^{\alpha\beta}_t=\alpha\left(\dfrac{\beta}{\alpha}\right)_x\! u+\left(\dfrac{\beta}{\alpha}\right)_tv^{\alpha}.
\end{equation}
Consider the system
\begin{equation}\label{potsysLinun}
v^\alpha_x=\alpha u,\quad v^\alpha_t=\alpha u_x-\alpha_xu,\quad
v^\beta_x=\beta u,\quad v^\beta_t=\beta u_x-\beta_xu
\end{equation}
that is the union of two potential systems of form~\eqref{potsysLin} corresponding to the local conservation laws
with the conserved vectors $(F^{\alpha},G^{\alpha})$ and $(F^{\beta},G^{\beta})$.
In a similar way to the previous cases we can state that
the second-level potential conservation law with conserved vector~\eqref{dvcllheAB} is trivial
on the solution manifold of system~\eqref{potsysLinun}
since
\[F^{\alpha\beta}=D_x\left(\dfrac{\beta}{\alpha} v^{\alpha}-v^{\beta}\right) \quad\mbox{and}
\quad G^{\alpha\beta}=-D_t\left(\dfrac{\beta}{\alpha} v^{\alpha}-v^{\beta}\right).
\]
Moreover, systems~\eqref{potsysLinsec} and~\eqref{potsysLinun} are connected via the local substitution
\[
w^{\alpha\beta}=\dfrac{\beta}{\alpha} v^{\alpha}-v^{\beta}.
\]
It implies that really system~\eqref{potsysLinun}
is generated by only three independent equations.
We can choose e.g. the equations
\[v^\alpha_x=\alpha u, \quad v^\beta_x=\beta u, \quad
\dfrac{\beta}{\alpha} v^{\alpha}_t-v^{\beta}_t=
\alpha\left(\dfrac{\beta}{\alpha}\right)_x\! u.
\]

\medskip

As a result of our analysis, we can formulate the following statement.

\begin{theorem}\label{Theorem2LevelPotSystems}
For any non-linearized equation~\eqref{eqf1} and the linear heat equation
the potential systems of the second level, which are constructed by means of
using single conservation law of the simplest potential systems, are equivalent to
first-level potential systems obtained with pairs of conservation laws.
\end{theorem}

\section{Potential conservation laws of linear heat equation}\label{SectionPotConsLawLHE}

With respect to $G^{\Equiv}$-equivalence the linear heat equation is the unique linear equation in
class~\eqref{eqf1}.
Investigation of its general potential conservation laws plays the major role in classification of
potential conservation laws for linearizable equations in class~\eqref{eqf1}
and, therefore, for whole class~\eqref{eqf1}.
(The simplest potential conservation laws are studied in the previous section.)

As proved in Theorem~\ref{TheoremCLCLofDCEs},
the linear heat equation has the infinite series of local conservation laws.
Fixing an arbitrary $p\in\mathbb{N}$ and
choosing $p$~linear independent solutions $\bar\alpha=(\alpha^1,\ldots,\alpha^p)$ of
the backward linear heat equation, we obtain $p$~linear independent conservation laws
with the conserved vectors $(F^s,G^s)=(\alpha^s u,\alpha^s_xu-\alpha^s u_x)$.
(Hereafter $s=\overline{1,p}$.)
In view of Theorem~\ref{TheoremOnIndepPotOfEvolEqs}
the potentials $\bar v=(v^1,\ldots,v^p)$ introduced with these conservation laws by the formulas
\begin{equation}\label{potsyslinP}
v^s_x=\alpha^s u,\quad v^s_t=\alpha^s u_x-\alpha^s_xu
\end{equation}
are independent in the sense of Definition~\ref{DefinitionOfPotentialDependence}.

For the linear heat equation the complete set of first level potential conservation laws
is indeed the union set of conservation laws of systems~\eqref{potsyslinP}
corresponding to all possible values of~$p$ and $p$-tuples~$\bar\alpha$.
The following theorem is true.

\begin{theorem}\label{TheoremOnPotConsLawsOfLHE}
Any local conservation law of system~\eqref{potsyslinP} is equivalent
on the manifold of system~\eqref{potsyslinP} to a local conservation law of the linear heat equation.
\end{theorem}

\begin{corollary}
For the linear heat equation potential conservation laws of any level are equivalent to local ones
on the manifolds of the corresponding potential systems,
and potentials of any level are locally expressed via local variables $t$, $x$, $u_{(r)}$ (for some $r$)
and potentials of the first level~only.
\end{corollary}

We present the proof of Theorem~\ref{TheoremOnPotConsLawsOfLHE} as the chain of simple and nice lemmas.

\begin{lemma}\label{LemmaOnLocConsLawsOfPotSystemForLHE}
Any local conservation law of system~\eqref{potsyslinP} is equivalent to that with the conserved vector
$(Ku,K_xu-Ku_x)$ where the function $K=K(t,x,\bar v)$ is determined by the system
\begin{equation}\label{SystemOnK}
K_t+K_{xx}=0, \qquad \alpha^sK_{xv^s}-\alpha^s_xK_{v^s}=0.
\end{equation}
\end{lemma}

\begin{proof}
Consider a local conservation law of system~\eqref{potsyslinP} in the most general form,
where the conserved vector is a vector-function of $t$, $x$ and  derivatives of the functions $u$ and $v^s$
from the zero order up to some finite one.
Taking into account system~\eqref{potsyslinP} and its differential consequences, we can exclude
dependence of the conserved vector on the all (non-zero order) derivatives of $v^s$ and the derivatives of~$u$
containing differentiation with respect to~$t$.
Similarly to Lemma~\ref{LemmaOnOrderOfConsLawsOfDCEs} we can prove that
the reduced conserved vector~$(F,G)$ does not depend on (non-zero order) derivatives of~$u$ and, moreover,
$F=F(t,x,\bar v)$, $G=-\alpha^sF_{v^s}(t,x,\bar v)u+G^0(t,x,\bar v)$. The function~$F$ and $G^0$
satisfy the system
\[
\alpha^s\alpha^{s'}F_{v^sv^{s'}}=0, \qquad
\alpha^sG^0_{v^s}=2\alpha^s_xF_{v^s}+\alpha^sF_{xv^s}, \qquad
F_t+G^0_x=0.
\]
Let us pass on to the equivalent conserved vector $(\widetilde F,\widetilde G)$, where
$\widetilde F=F+D_xH$, $\widetilde G=G-D_tH$ and $H=H(t,x,\bar v)$ is a solution of the equations
$H_x=-F$, $H_t=G$. (The variables $v^s$ is assumed as parameters in the latter equations.) Then
$\widetilde F=Ku$, $\widetilde G=K_xu-Ku_x$. The function $K=\alpha^sH_{v^s}$ depends on $t$, $x$ and $\bar v$
and satisfy system~\eqref{SystemOnK}.
\end{proof}

\begin{lemma}\label{LemmaOnDiffConsOfPotSystemForLHE}
Let the solutions~$\alpha^s=\alpha^s(t,x)$ and $\beta^s=\beta^s(t,x)$ of the (backward) linear heat equation
satisfy the additional condition $\alpha^s_x\beta^s-\alpha^s\beta^s_x=0$.
Then for any $i,j\in{\mathbb N}$
\begin{equation}\label{EqAlphaBetaIJ}
\alpha^s_i\beta^s_j-\alpha^s_j\beta^s_i=0.
\end{equation}
Hereafter the subscripts $i$ and $j$ denote the $i$-th and $j$-th order derivatives with respect to $x$.
\end{lemma}

\begin{proof}
We make the proof by means of mathematical induction with respect to the value~$i+j$.

Equation~\eqref{EqAlphaBetaIJ} is trivial for $i+j=0$, coincides with the additional condition for $i+j=1$ and
is obtained from this condition by means of differentiation with respect to $x$ if $i+j=2$.

Let us suppose that the Lemma's statement is true if $i+j=m-1$ and $i+j=m$
and prove it for $i+j=m+1$. Acting on equation~\eqref{EqAlphaBetaIJ} where $i+j=m-1$ with the operator $\p_t+\p_{xx}$
and taking into account the conditions $\alpha^s_t+\alpha^s_{xx}=0$ and $\beta^s_t+\beta^s_{xx}=0$,
we obtain the equation \[\alpha^s_{i+1}\beta^s_{j+1}-\alpha^s_{j+1}\beta^s_{i+1}=0.\]
Therefore, the statement is true for $i'+j'=m+1$, $1\le i',j'\le m$ ($i'=i+1$, $j'=j+1$).
It~remains to perform the proof in the case $i'=m+1$, $j'=0$ (or equivalently $i'=0$, $j'=m+1$).
For these values of $i'$ and $j'$ the statement is produced
by subtracting the induced above equation $\alpha^s_m\beta^s_1-\alpha^s_1\beta^s_m=0$ from the results of
differentiation of equation~\eqref{EqAlphaBetaIJ} where $i=m$, $j=0$ with respect to $x$.
\end{proof}

Let $W(\varphi^1,\ldots,\varphi^l)$ denotes the Wronskian of the functions $\varphi^1$, \ldots, $\varphi^l$
with respect to the variable~$x$, i.e. $W(\varphi^1,\ldots,\varphi^l)=\det(\varphi^j_i)_{i,j=1}^{\;l}$.

\begin{lemma}\label{LemmaOnlinearDependenceOfSolutionsOfLHE}
The solutions~$\varphi^1=\varphi^1(t,x)$, \ldots, $\varphi^l=\varphi^l(t,x)$ of
a linear evolution equation $L\varphi=0$ are linear dependent iff $W(\varphi^1,\ldots,\varphi^l)=0$.
\end{lemma}

\begin{proof}
Since the equation $L\varphi=0$ is linear and evolution the operator $L$ is the sum of $\p_t$ and
linear differential operator with respect to $x$ with the coefficients depending on $t$ and $x$.
If the functions $\varphi^1$, \ldots, $\varphi^l$ are linear dependent then
the equality $W(\varphi^1,\ldots,\varphi^l)=0$ is obvious.
Let us prove the inverse statement.

In the case $l=2$ the condition $W(\varphi^1,\varphi^2)=0$ implies $\varphi^2=C\varphi^1$,
where $C$ is a smooth function of~$t$. Acting on the latter equality with the operator $L$,
we obtain $C_t\varphi^1=0$, i.e. $C=\const$ or $\varphi^1=0$.
In any case the functions~$\varphi^1$ and~$\varphi^2$ are linear dependent.

Suppose $W(\varphi^1,\ldots,\varphi^l)=0$.
Without loss of generality we can assume $W(\varphi^1,\ldots,\varphi^{l-1})\ne 0$.
(Otherwise we consider a less value of $l$.) Then
$\varphi^l=C^k\varphi^k,$
where $C^k$ are smooth functions of $t$ and the superscript $k$ runs from 1 to $l-1$.
Action of the operator $L$ on the latter equality  results in the equation $C^k_t\varphi^k=0$
that implies, in view of the condition $W(\varphi^1,\ldots,\varphi^{l-1})\ne 0$,
$C^k=\const$. It gives the Lemma's statement.
\end{proof}

\begin{lemma}\label{LemmaOnWronskianForLHE}
If $\alpha^s_i\beta^s_j-\alpha^s_j\beta^s_i=0$ for $0\le i<j\le p$
then $W(\alpha^1,\ldots,\alpha^p,\beta^{s'})=0$ for any $s'$.
\end{lemma}

\begin{proof}
Let $M^{s'}_{ij}$ denote the $(p-1)$-th order minor of $W(\bar\alpha,\beta^{s'})$,
which is obtained by means of deletion of $s'$-th and $(p+1)$-th columns
corresponding to the functions~$\alpha^{s'}$ and~$\beta^{s'}$ and $i$-th and $j$-th rows.
Let us multiply the equation~$\alpha^s_i\beta^s_j-\alpha^s_j\beta^s_i=0$ by
$(-1)^{i+j+s'+p+1}M^{s'}_{ij}$ and convolve with respect to the indices~$i$ and~$j$.
In view of the Laplace theorem on determinant expansion we obtain
\[W(\bar\alpha|_{\alpha^s\rightsquigarrow\alpha^{s'}},\beta^{s})=0.\]
Here the sign~``$\rightsquigarrow$'' means that
the function~$\alpha^s$ is substituted instead of the function~$\alpha^{s'}$
and we have summation over the index~$s$.
The Lemma's statement easily follows from the latter equation since for any fixed $s\ne s'$ we have
$W(\bar\alpha|_{\alpha^s\rightsquigarrow\alpha^{s'}},\beta^{s})=0$.
\end{proof}

\begin{lemma}\label{LemmaGenSolutionOfSystemOnK}
The general solution of system~\eqref{SystemOnK} can be presented in the form
\begin{equation}\label{ExpressionForK}
K=\alpha^sH_{v^s}+\beta^0,
\end{equation}
where $H$ is an arbitrary smooth function of $\bar v$,
$\beta^0=\beta^0(t,x)$ is an arbitrary solution of the backward linear heat equation.
\end{lemma}

\pagebreak

\begin{proof}
In view of Lemma~\ref{LemmaOnLocConsLawsOfPotSystemForLHE}
the functions~$\alpha^s$ and $\beta^s=K_{v^s}$ satisfy the conditions of Lemma~\ref{LemmaOnDiffConsOfPotSystemForLHE}
and, therefore, the ones of Lemma~\ref{LemmaOnWronskianForLHE},
and the variables~$\bar v$ are assumed as parameters. Since $\alpha^s$ are linear independent
it implies $K_{v^{\sigma}}=C^{\sigma s}\alpha^s$, where $C^{\sigma s}$
are smooth functions of the variables~$\bar v$ only.
Hereafter the indices~$s$, $\sigma$ and $\varsigma$ run from~1 to $p$.
The expressions for the cross derivatives
$K_{v^{\sigma}v^{\varsigma}}=C^{\sigma s}_{v^{\varsigma}}\alpha^s=C^{\varsigma s}_{v^{\sigma}}\alpha^s$
result in the equation $C^{\sigma s}_{v^{\varsigma}}=C^{\varsigma s}_{v^{\sigma}}$ which can be easily integrated:
$C^{\sigma s}=P^s_{v^{\sigma}}$ for some smooth function~$P^s$ of the variables~$\bar v$.
Substituting the expressions for~$C^{\sigma s}$ in the equations on $K$ and integrating, we obtain
$K=\alpha^sP^s+\beta^0$, where $\beta^0=\beta^0(t,x)$ is a solution of the backward heat equation.
The latter equality and the equation $\alpha^sK_{xv^s}-\alpha^s_xK_{v^s}=0$ together imply
the equation
$(\alpha^{\sigma}_x\alpha^{\varsigma}-\alpha^{\sigma}\alpha^{\varsigma}_x)
(P^{\varsigma}_{v^{\sigma}}-P^{\sigma}_{v^{\varsigma}})=0$.
Analogously to Lemma~\ref{LemmaOnWronskianForLHE} we can state for any $i,j\in{\mathbb N}$
\begin{equation}\label{EqAlphaWij}
(\alpha^{\sigma}_i\alpha^{\varsigma}_j-\alpha^{\sigma}_j\alpha^{\varsigma}_i)
(P^{\varsigma}_{v^{\sigma}}-P^{\sigma}_{v^{\varsigma}})=0.
\end{equation}

Let $M^{\sigma'\varsigma'}_{ij}$ denote the $(p-2)$-th order minor of $W(\bar\alpha)$,
which is obtained by means of deletion of $\sigma'$-th and $\varsigma'$-th columns
corresponding to the functions~$\alpha^{\sigma'}$ and~$\alpha^{\varsigma'}$ and $i$-th and $j$-th rows.
Let us multiply the equation~\eqref{EqAlphaWij} by $(-1)^{i+j+\sigma'+\varsigma'}M^{\sigma'\varsigma'}_{ij}$
and convolve with respect to the indices~$i$ and~$j$.
In view of the Laplace theorem on determinant expansion we obtain
\begin{equation}\label{EqAlphaWij1}
W\bigl(\bar\alpha|_{
\alpha^{\sigma}\rightsquigarrow\alpha^{\sigma'}\!\!,\;
\alpha^{\varsigma}\rightsquigarrow\alpha^{\varsigma'}}\!
\bigr)
(P^{\varsigma}_{v^{\sigma}}-P^{\sigma}_{v^{\varsigma}})=0.
\end{equation}
Here the sign~``$\rightsquigarrow$'' means that
the functions~$\alpha^{\sigma}$ and $\alpha^{\varsigma}$ is substituted
instead of the function~$\alpha^{\sigma'}$ and $\alpha^{\varsigma}$ correspondingly
and we have summation over the indices~$\sigma$ and $\varsigma$.
For any fixed $\sigma\ne \sigma'$ and $\varsigma\ne \varsigma'$ we have
$W\bigl(\bar\alpha|_{
\alpha^{\sigma}\rightsquigarrow\alpha^{\sigma'}\!\!,\;
\alpha^{\varsigma}\rightsquigarrow\alpha^{\varsigma'}}\!
\bigr)=0$.
Since $W(\bar\alpha)\ne 0$ in view of linear independence of the functions~$\alpha^s$,
equation~\eqref{EqAlphaWij1} implies $P^{\varsigma'}_{v^{\sigma'}}-P^{\sigma'}_{v^{\varsigma'}}=0$,
i.e. $P^s=H_{v^s}$ for some smooth function~$H$ of~$\bar v$.
\end{proof}

In view of Lemma~\ref{LemmaGenSolutionOfSystemOnK} the conserved vector $(Ku,K_xu-Ku_x)$
from Lemma~\ref{LemmaOnLocConsLawsOfPotSystemForLHE} has the form
$(\beta^0u+D_xH,\beta^0_xu-\beta^0u_x-D_tH)$.

The proof of Theorem~\ref{TheoremOnPotConsLawsOfLHE} is completed.

\section{Potential conservation laws of\\ nonlinear diffusion--convection equations}\label{SectionGenPotConsLaws}

More general potential conservation laws than simplest ones are admissible only if the investigated system has
\begin{itemize}\itemsep=0ex
\item
either more than one linear independent local conservation laws (and, therefore,
we can introduce a number of different potentials for the first iteration step)
\item
or non-trivial simplest potential conservation laws.
\end{itemize}
As shown in Section~\ref{SectionSimplestPotentialConservationLaws}, it is possible in class~\eqref{eqf1}
only for special values of the parameter functions~$A$ and $B$.
In view of results of Section~\ref{SectionSimplestPotentialConservationLaws} for the linearizable equations and
Theorem~\ref{TheoremOnPotConsLawsOfLHE}, we can formulate the following statement.

\begin{theorem}
For any linearizable equation from class~\eqref{eqf1} all potential conservation laws of the second level are equivalent
on the manifold of the corresponding potential systems to potential conservation laws of the first level.
\end{theorem}

To investigate completely the potential conservation laws of equations from class~\eqref{eqf1},
it remains to study the subclasses with $B=0$ and $B=A$, equations from which have two independent
local conservation laws, and the subclass $B=\int\! A+uA$ reduced to $B=A$
by means of potential equivalence transformations.

\begin{theorem}\label{TheoremGenPotConsLawB0BA}
All potential conservation laws of any equation from class~\eqref{eqf1} with $B=0$ or $B=A$ are trivial
on the manifold of the united potential
systems~\eqref{potsysB0gen}, \eqref{potsysB0spec} and~\eqref{potsysBAgen}, \eqref{potsysBAspec}
constructed with pairs of independent local conservation laws.
\end{theorem}

\begin{proof}
Consider the united system \eqref{potsysB0gen}, \eqref{potsysB0spec} (or~\eqref{potsysBAgen}, \eqref{potsysBAspec}).
(Below we write down the differences of the second case with the first one in brackets.)
Similarly to Lemma~\ref{LemmaRedConsLaws}, we can assume without loss of generality that
$F=F(t,x,v^1,v^2)$ and $G=G(t,x,u,v^1,v^2)$. Let us split the equation $D_tF+D_xG=0$
on the manifold determining by the united system.
Integration of one from the obtained equations results in the following expression for the flux~$G$:
$G=-(QF)\textstyle\int\!A+G^0,$ where $G^0=G^0(t,x,v^1,v^2)$ and $Q=\p_{v^1}+x\p_{v^2}$ ($Q=\p_{v^1}+e^x\p_{v^2}$).
The other equations form the system on the functions~$F$ and~$G^0$:
\[
Q^2F=0, \quad QG^0=0,\quad F_t+G^0_x=0,\quad (QF)_x+F_{v^2}=0 \quad (\,(QF)_x-F_{v^1}=0\,).
\]
Therefore, $F=F^1v^1+F^0$, and $F^1$, $F^0$ and $G^0$ are functions of $t$, $y=x$ and $\omega=xv^1-v^2$
($\omega=e^xv^1-v^2$) for which $F^0_{\omega}=F^1_y$, $F^1_t=-G^0_{\omega}$, $F^0_t=-G^0_y$.
The latter system implies existence of such function~$H=H(t,y,\omega)$ that
\[
F^1=H_\omega\quad (F^1=e^xH_\omega),\quad F^0=H_y,\quad G^0=-H_t.
\]
Then, $F=D_xH$, $G=-D_tH$, i.e. the conservation law is trivial.
\end{proof}

As shown above, there exists the following chain of local transformations between potential systems:
1.3 $\longleftrightarrow$ 1.2 $\longleftrightarrow$ \eqref{potsysBAgen}, \eqref{potsysBAspec},
i.e. system 1.3 is locally equivalent to system~\eqref{potsysBAgen}, \eqref{potsysBAspec}.
In view of this fact and Theorem~\ref{TheoremGenPotConsLawB0BA} we obtain the following statement.

\begin{theorem}
On the manifold of the potential system~1.3
all potential conservation laws of any equation from class~\eqref{eqf1} with $B=\int\!A+uA$ are trivial.
\end{theorem}

Summarizing the above results, we note that up to $G^{\Equiv}$-equivalence
the hierarchy of  conservation laws (including local ones) for diffusion--convection equations~\eqref{eqf1} has the form:
\begin{itemize}\itemsep=0ex
\item
the ``common'' local conservation law (Case~1) for arbitrary values of the parameter functions~$A$ and~$B$;
\item
two independent local conservation laws if $B=0$ (Cases~1 and~2) or $B=A$ (Cases~1 and~3);
\item
one ``common'' local conservation law (Case~1) and one simplest potential that (Case~1.3) if $B=\int\!A+uA$;
\item
the infinite series of local conservation laws (Case~4) for the linear heat equation;
\item
one ``common'' local conservation law (Case~1) and
the infinite series of simplest potential conservation laws (Case~1.6) for the Burgers equation;
\item
two independent local conservation laws for the $u^{-2}$-diffusion equation (Cases~1 and~2) and
the equation $u_t=(u^{-2}u_x)_x+u^{-2}u_x$ (Cases~1 and~3) as subcases of $B=0$ and $B=A$
and the infinite series of simplest potential conservation laws (Cases~1.4 and~1.5) additionally.
\end{itemize}

\begin{note}
Above we did not consider in an explicit form action of transformations from Lie symmetry groups
on conservation laws of corresponding equations or potential systems.
For the majority of cases this action is quite trivial.
For example, we use translations with respect to~$x$ to normalize the constant $\varepsilon$ in Case~3.
In Case~2 the same translations result in adding the ``common'' conservation law
to the special one of this case.

A non-obvious connection between independent conservation laws can be established
only for~$A=u^{-4/3}$, $B=0$ (Case~1 and Case~2) by means of the transformation
$\tilde t=t$, $\tilde x=1-x^{-1}$, $\tilde u=x^3u$ from the Lie symmetry group of the corresponding equation.
This fact was first discovered in~\cite{Kara&Mahomed2002} in the framework of the ``operator'' approach.
It should be mentioned that the values $A=u^{-4/3}$, $B=0$ give rise to the equation
which distinguishes from non-linear diffusion--convection equations~\eqref{eqf1} by singular Lie symmetry properties.

The Lie symmetry group~${\mathcal G}_{-}$ of the linear heat equation contains
infinite dimensional normal subgroup~${\mathcal G}_{-}^0$ formed by
the linear superposition transformations $\tilde u=u+f(t,x)$,
where $f=f(t,x)$ is an arbitrary solution of the same equation.
Up to the equivalence relation, transformations from~${\mathcal G}_{-}^0$
act identically on the set of conservations laws of Case~4.
Action of the finite dimensional factor group~${\mathcal G}_{-}/{\mathcal G}_{-}^0$ on this set induces
the analogous factor group~${\mathcal G}_{+}/{\mathcal G}_{+}^0$ on the set of solutions of
the backward linear heat equation, which is varied over by the parameter-function~$\alpha$.
\end{note}

The hierarchy of conservation laws generates the complete set of locally inequivalent
potential systems for the class under consideration:
\begin{itemize}\itemsep=0ex
\item
``common'' potential system~\eqref{PotSysGen} (Case~1);
\item
additional simplest potential systems~\eqref{potsysB0spec} (Case~2) or~\eqref{potsysBAspec} (Case~3)
if $B=0$ or $B=A$ correspondingly;
\item
second level potential systems~\eqref{potsys2B0gen} (Case~1.1) and~\eqref{potsys2BAgen} (Case~1.2)
(which are really equivalent to the united potential systems of the first level)
if $B=0$ or $B=A$ correspondingly;
\item
system~\eqref{potsyslinP} with arbitrary number of locally independent potentials for
the linear heat equation.
\end{itemize}

Potential symmetries arising for equations~(\ref{eqf1}) from Cases~1 and~1.1 of Table~1 were studied
by C.~Sophocleous~\cite{Sophocleous1996}.
Complete investigation of the potential system~\eqref{PotSysGen} (Case~1) with the symmetry point of view
was carried out in~\cite{Popovych&Ivanova2005PETs}.

\section{Conclusion}

The notions and methods proposed in the paper are simple and effective tools for investigation of
both local and pure potential conservation laws. They can be applied to a wide range of physically interesting
systems of differential equations. At the same time, there exist a number of unresolved problems, in particular,
on determining the number of necessary iterations for construction of an exhaustive list of independent potential
conservation laws or on connections of our framework
with Wahlquist--Estabrook prolongation structures~\cite{Wahlquist&Estabrook1975}.

The adduced results for diffusion--convection equations can be developed and generalized in a number of directions.
So, studying different kinds of symmetries (Lie, nonclassical, generalized ones) of constructed potential systems,
we may obtain the corresponding kinds of potential symmetries
(usual potential, nonclassical potential, generalized potential ones).
Let us note that investigation of generalized symmetries is natural for potential systems,
since potentials introduced with equivalent conservation laws are related, in general, via transformations
depending on derivatives of local dependent variables. Analogously, local equivalence transformations between
potential systems constructed for different initial equations result in nonlocal (potential)
equivalence transformations for the class under consideration. In such way it is possible to find new connections
between well-studied diffusion--convection equations~\cite{Popovych&Ivanova2005PETs}.
We believe that the same approach can be used for investigation of wider classes of differential equations,
e.g. variable coefficient diffusion--convection equations.
We also plan to study conservation laws of more general structure (e.g. ones with pseudopotentials).

\subsection*{Acknowledgements}

The authors are grateful to
 V.~Boyko, A.~Nikitin, A.~Sergyeyev, I.~Yehorchenko and A.~Zhalij
for useful discussions and interesting comments.
Research of NMI was supported by National Academy of Science of Ukraine
in the form of the grant for young scientists.
ROP thanks Prof.~F.~Ardalan and Dr.~H.~Eshraghi
(School of Physics, Institute for Studies in Theoretical Physics and Mathematics, Tehran, Iran)
for hospitality and support during writing this paper.

\end{document}